\documentclass[showpacs, amsmath, amsfonts, amssymb, aps, superscriptaddress]{revtex4}

\usepackage{appendix} 
\usepackage{tabularx}
\usepackage[dvipsnames]{xcolor}
\usepackage{comment}
\usepackage{enumitem}
\usepackage{graphicx}
\usepackage{float}
\usepackage{dcolumn}
\usepackage{bm}
\usepackage{mathrsfs}
\usepackage{amsmath}

\usepackage{amsfonts}
\usepackage{dsfont}
\usepackage{graphicx}
\usepackage{subcaption}

\begin{document}

\title{Exact Spinning Morris–Thorne Wormhole: Causal Structure, Shadows, and Multipole Moments}
\author{Davide Batic}
\email{davide.batic@ku.ac.ae (corresponding author)}
\affiliation{Department of Mathematics, Khalifa University of Science and Technology, PO Box 127788, Abu Dhabi, United Arab Emirates}

\author{Denys Dutykh}
\email{denys.dutykh@ku.ac.ae}
\affiliation{Department of Mathematics, Khalifa University of Science and Technology, PO Box 127788, Abu Dhabi, United Arab Emirates}

\author{Mark Essa Sukaiti}
\email{100064482@ku.ac.ae}
\affiliation{Department of Mathematics, Khalifa University of Science and Technology, PO Box 127788, Abu Dhabi, United Arab Emirates}
\date{\today}

\begin{abstract}
We construct an exact spinning generalisation of the Morris–Thorne traversable wormhole supported by an anisotropic fluid. Within the Teo wormhole ansatz with unit lapse and Morris–Thorne shape function, we solve analytically for the frame–dragging function and obtain a two–parameter family of asymptotically flat solutions labelled by the throat radius $r_0$ and total angular momentum $J$. Curvature scalars and stress–energy components are given in closed form, showing a regular throat, equatorial reflection symmetry, and violations of all standard energy conditions, as required for traversable wormholes. We analyse the causal structure and show that, despite the presence of an ergoregion for sufficiently large $|J|$, the coordinate time defines a global temporal function, so the spacetime is stably causal and free of closed timelike curves. The optical appearance is studied via photon trajectories. The resulting shadows are smaller than Kerr’s and depend on the wormhole shape. Finally, we compute the Geroch–Hansen multipole moments and find a massless but spinning configuration with distinctive higher multipoles that encode the throat scale.

\end{abstract}

\maketitle

\section{Introduction}
Traversable wormholes continue to provide a useful theoretical laboratory for probing the interplay between geometry, topology, and exotic matter in general relativity. Since the seminal work of Morris and Thorne \cite{Morris1988AJP} and the Ellis–Bronnikov constructions \cite{Bronnikov1973APP, Ellis1973JMP, Ellis1974JMP, Ellis1979GRG}, it has been clear that static, spherically symmetric traversable wormholes require violations of the usual energy conditions, typically implemented via phantom scalar fields or anisotropic fluids tuned at the throat. These configurations have been analysed in detail at the level of their local geometry, global structure, and matter content, and they form a standard baseline for assessing more general wormhole spacetimes.

From an astrophysical point of view, if wormholes are to be considered even as effective models, one must understand their rotating counterparts. Rotation introduces qualitatively new features such as frame dragging, ergoregions, possible superradiant processes, and a richer causal structure, including the potential for closed timelike curves. It also changes observational signatures, from gravitational lensing and shadows to orbital dynamics and gravitational‑wave emission. A controlled family of rotating wormhole solutions, with explicit matter sources and asymptotically flat ends, is therefore essential for connecting wormhole physics to current and future observations.

Most rotating‑wormhole constructions to date rely on slow‑rotation expansions or on numerical integration \cite{Kashargin2008GC, Kashargin2008PRD, Kuhfittig2003PRD, Azad2023, Azad2024PLB, Kleihaus2014PRD, Hoffmann2018PRD, Chew2019PRD}. Nonetheless, analytic stationary solutions do exist in specific settings, \emph{e.g.} via Ehlers/Harrison transformations in Einstein–Maxwell theory, in overcharged Kerr--Newman--NUT geometries, and for particular matter sources such as Casimir stresses or three‑form fields \cite{Cisterna2023PRD, Clement2023PLB, Tanghpati2024NPB}. For the matter models most frequently considered in the wormhole literature, such as phantom scalars and anisotropic fluids with well‑controlled throat and gauge conditions, rotating solutions are, apart from the aforementioned special analytic cases, almost entirely limited to slow‑rotation expansions and numerical studies. The present work occupies precisely this analytically tractable sector. Our aim is to construct and analyse a fully explicit rotating generalisation of the Morris–Thorne wormhole, supported by an anisotropic fluid, and to characterise its causal and observational properties in detail.

On the phenomenological side, rotating wormholes have been investigated as potential alternatives to Kerr black holes in shadow imaging and strong‑field lensing. Teo’s metric provides a convenient stationary, axisymmetric ansatz with a prescribed rotation function and throat geometry \cite{Teo1998PRD}, which has been used extensively in studies of photon motion and shadows \cite{Deligianni2021PRD, Deligianni2021PRDa, Shaikh2018PRD}. However, in many of these works the matter content is left implicit or treated effectively, the rotation profile is chosen ad hoc, and the connection to standard Morris–Thorne–type throats is not explicit. Moreover, the dependence of observable quantities, such as the shadow size and shape, on the underlying wormhole shape function remains unclear, and the causal structure, in particular, the compatibility of ergoregions with acceptable causality, is often assumed rather than derived.

A complementary, and more invariant, way to characterise a rotating spacetime is through its multipole structure near spatial infinity. The Geroch--Hansen (GH) multipole moments \cite{Geroch1970JMP, Geroch1970JMPa, Hansen1974JMP} provide a coordinate‑independent description of the far field for stationary, asymptotically flat spacetimes and play a central role in black‑hole uniqueness theorems and no‑hair results \cite{Thorne1980RMP, Gursel1983GRG, Mayerson2023SP, Heusler2010}. For black holes, the GH multipoles are tightly constrained. For wormholes, by contrast, they are largely unconstrained and can, in principle, encode nontrivial information about the throat and the matter supporting it. Yet explicit GH multipole computations for rotating wormholes with a clear matter model are scarce, especially beyond the lowest moments.

In this paper, we address these gaps by constructing and analysing an exact spinning Morris--Thorne wormhole within the Teo ansatz. Starting from the static Morris--Thorne/Ellis--Bronnikov geometry with shape function $b(r)=r_0^2/r$ and anisotropic fluid source, we adopt a Teo‑like rotating extension with unit lapse and spherical throat geometry, and we impose that the matter content remains an anisotropic fluid in an orthonormal comoving frame. The Einstein equations, together with consistency conditions on the anisotropic pressures, reduce the rotation function to satisfy a single ordinary differential equation in the proper radial distance $\ell$, which we solve in closed form. This yields a two‑parameter family of asymptotically flat spinning wormholes labelled by the throat radius $r_0$ and the total angular momentum $J$, with the metric given explicitly in \eqref{metricell}.

Our analysis is organised around three main questions. Does the rotating deformation remain regular at the throat, and how does it modify the energy‑condition violations already present in the static Morris--Thorne wormhole? In that regard, we compute curvature invariants and the stress–energy components in closed form, showing manifest regularity at $\ell=0$, equatorial reflection symmetry, and violations of all standard energy conditions, with rotation enhancing the exotic character off the axis. A Komar integral associated with the axial Killing vector confirms that $J$ coincides with the total angular momentum of the spacetime. The second question we address is how rotation affects the causal properties and optical appearance of the wormhole. We show that although there exists a sharp condition on $|J|$ that controls the onset of an ergoregion, the geometric background still admits a global time function and remains stably causal, with no closed timelike curves. Building on \cite{Deligianni2021PRD, Shaikh2018PRD}, we compute the corresponding shadows and compare them with those of a Kerr black hole, finding systematically smaller shadows and a nontrivial dependence on the wormhole shape function, in contrast with some earlier claims. Finally, what is the GH multipole content of this spinning wormhole, and how does it differ from Kerr? By recasting the metric in Weyl–Lewis–Papapetrou form and constructing the Ernst potential with the appropriate improved twist \cite{Mayerson2023SP, Fodor1989JMP, Sotiriou2004CQG}, we compute the GH moments and show that the configuration is massless but spinning, with vanishing mass quadrupole and the first nontrivial structure appearing at the octupole level, where the throat radius leaves a clear imprint. This multipole pattern is qualitatively distinct from the Kerr pattern and provides a natural input for post‑Newtonian and EMRI modelling.

The rest of the paper is organised as follows. In Sec. II, we review the static Morris--Thorne wormhole with anisotropic fluid source and introduce the rotating Teo‑type ansatz. We derive the reduced field equations, solve exactly for the rotation function, and present the full spinning Morris–Thorne metric, together with curvature invariants and stress–energy components. In Sec. III, we turn to the main physical implications. More precisely,  Sec. III A discusses the shadows of the rotating wormhole and their comparison with Kerr, while Sec. III B and C analyse the ergoregion and the causal structure, for which we confirm the absence of closed timelike curves. Finally, Sec. III D presents the Geroch--Hansen multipole moments and contrasts the resulting multipole hierarchy with that of Kerr. Section IV summarises our findings and outlines prospects for applications to orbital dynamics, gravitational-wave phenomenology, and lensing.

\section{DERIVATION OF THE SPINNING MORRIS-THORNE WORMHOLE}

As the wormhole seed metric, \emph{i.e.}, in the absence of rotation, we consider a static, spherically symmetric spacetime. In natural units $c = G_N = 1$, the line element of such a wormhole can be written as \cite{Morris1988AJP, Ellis1973JMP, Ellis1979GRG, Ellis1974JMP, Bronnikov1973APP}
\begin{equation}\label{metric}
  ds^2 = -e^{2\Phi(r)}dt^2+\frac{dr^2}{1-\frac{b(r)}{r}} + \frac{r^2 d\chi^2}{1-\chi^2} + r^2(1-\chi^2)d\varphi^2, \quad \chi\in[-1,1], \quad \varphi\in[0,2\pi).
\end{equation}
Here, $\Phi$ and $b$ are the redshift and shape functions, respectively. We assume that the radial coordinate $r$ increases monotonically from its minimum value $r_0$, representing the throat of the wormhole, to spatial infinity. The matter content acting as a source of  the geometry described by \eqref{metric} is modelled in terms of an anisotropic fluid, with energy-momentum tensor given by
\begin{equation}\label{emt}
 T^\alpha{}_\beta=(\rho+p_t)u^\alpha u_\beta+p_t\delta^\alpha{}_\beta+(p_r-p_t)\ell^\alpha\ell_\beta,    
\end{equation}
where $\ell^\alpha$ is a unit spacelike vector orthogonal to the fluid four-velocity $u^\alpha$, that is $\ell^\alpha \ell_\alpha=1$ and $\ell^\alpha u_\alpha=0$. Moreover, $u^\alpha$ must satisfy the condition $g_{\alpha\beta}u^\alpha u^\beta=-1$. These constraints require that
\begin{equation}\label{utnorot}
    u^\alpha=e^{-\Phi(r)}\delta^\alpha{}_t,\quad
    \ell^\alpha=\sqrt{1-\frac{b(r)}{r}}\delta^\alpha{}_r.
\end{equation}
Hence, the mixed energy-momentum tensor is $T^\alpha{}_\beta=\text{diag}(-\rho(r), p_r(r), p_t(r), p_t(r))$, where $\rho$ is the energy density, $p_r$ the radial pressure, and $p_t$ the tangential pressure measured orthogonally to the radial direction. By applying the Einstein field equations $G_{\alpha\beta}=8\pi T_{\alpha\beta}$ alongside the conservation equation $\nabla_\alpha T^{\alpha\beta}=0$, we obtain the following system of equations, where an  overdot denotes differentiation with respect to the radial coordinate
\begin{eqnarray}
&&\dot{b}-8\pi r^2\rho=0,\label{eq1}\\
&&2r(r-b)\dot{\Phi}-b-8\pi r^3 p_r=0,\label{eq2}\\
&&r^2(r-b)\ddot{\Phi}+\left[r(r-b)\dot{\Phi}+\frac{b-r\dot{b}}{2}\right](1+r\dot{\Phi})-8\pi r^3 p_t=0,\label{eq3}\\
&&r(\rho+p_r)\dot{\Phi}+r\dot{p}_r+2(p_r-p_t)=0.\label{eq4}
\end{eqnarray}
Using equations \eqref{eq1}--\eqref{eq3}, we can express the energy density, the radial and tangential pressures in terms of the redshift and shape functions as follows
\begin{eqnarray}
\rho&=&\frac{\dot{b}}{8\pi r^2},\label{ed}\\
p_r&=&\frac{1}{8\pi}\left[\frac{2}{r}\left(1-\frac{b}{r}\right)\dot{\Phi}-\frac{b}{r^3}\right],\label{pr}\\
p_t&=&\frac{1}{8\pi}\left(1-\frac{b}{r}\right)\left[\ddot{\Phi}+\dot{\Phi}^2-\frac{r\dot{b}-b}{2r(r-b)}\dot{\Phi}+\frac{\dot{\Phi}}{r}-\frac{r\dot{b}-b}{2r^2(r-b)}\right].\label{pt}
\end{eqnarray}
As a rotating extension of the static wormhole metric \eqref{metric}, we adopt a Teo-like ansatz \cite{Teo1998PRD} written in Boyer--Lindquist–type coordinates $(t,r,\chi,\varphi)$ with $\chi=\cos{\theta}$
\begin{equation}\label{rotmet}
  ds^2=-[N^2-(1-\chi^2)\kappa^2\omega^2]dt^2+\frac{r}{\Sigma}dr^2-2\omega(1-\chi^2)\kappa^2 dtd\varphi+\kappa^2\left[\frac{d\chi^2}{1-\chi^2}+(1-\chi^2)d\varphi^2\right]
\end{equation}
with $\Sigma=r-B$ and $\kappa=rK$. Here $N(r,\chi)$ is the lapse, $\omega(r,\chi)$ is the angular velocity of inertial frames, $B(r,\chi)$ generalises the shape function, and $K(r,\chi)$ encodes the departure of the $r=$const two-geometry from sphericity. The metric reduces to the static configuration \eqref{metric} in the non-rotating limit $J\to 0$, i.e.
\begin{equation}
N(r,\chi)\to e^{\Phi(r)},\qquad
B(r,\chi)\to b(r),\qquad
K(r,\chi)\to 1,\qquad
\omega(r,\chi)\to 0.
\end{equation}
Furthermore, as already observed by \cite{Teo1998PRD}, $J$ can be interpreted as the total angular momentum per unit mass of the wormhole if we impose the following asymptotic behaviour on $\omega$, that is
\begin{equation}\label{asympt}
\omega(r,\chi)=\frac{2J}{r^3}+\mathcal{O}\left(\frac{J^3}{r^7}\right).    
\end{equation}
In order to prevent misreadings, notice that Teo's $a$ appearing in \cite{Teo1998PRD} is the total angular momentum $J$, not the Kerr parameter $a =J/M$. Moreover, we model the matter content of the rotating wormhole as an anisotropic fluid. Let $\{U^{\alpha}, n_{(r)}^{\alpha}, n_{(\chi)}^{\alpha}, n_{(\varphi)}^{\alpha}\}$ be an orthonormal tetrad comoving with the fluid, with $U_{\alpha}U^{\alpha}=-1$, $n_{(i)\,\alpha}n_{(j)}^{\alpha}=\delta_{ij}$, and $U_{\alpha}n_{(i)}^{\alpha}=0$. In its mixed-index form, the energy–momentum tensor is
\begin{equation}\label{AEMT}
  \widetilde{T}^\alpha{}_\beta=\widetilde{\rho}U^\alpha U_\beta + P_r n^{\alpha}_{(r)}n_{(r)\beta} + P_\chi n^{\alpha}_{(\chi)}n_{(\chi)\beta} + P_\varphi n^{\alpha}_{(\varphi)}n_{(\varphi)\beta},    
\end{equation}
where $\widetilde{\rho}(r,\chi)$ is the energy density measured by comoving observers and $P_{r}(r,\chi), P_{\chi}(r,\chi), P_{\varphi}(r,\chi)$ are the principal pressures along the radial, polar, and azimuthal directions, respectively. The four-velocity is chosen to describe a ZAMO flow, \emph{i.e.} a congruence corotating with the local inertial frames. By definition, a ZAMO (Zero Angular Momentum Observer) has zero angular momentum, \emph{i.e.} $U_\varphi=g_{\varphi\varphi}(\Omega-\omega)U^t=0$, which implies that its angular velocity satisfies the condition $\Omega = \omega$. Accordingly, we take $U^{\alpha} = N^{-1}(1,0,0,\omega)$, with the normalization fixed by $U_{\alpha}U^{\alpha}=-1$.  Furthermore, if we impose $n_{(i)\,\alpha}n_{(j)}^{\alpha}=\delta_{ij}$, and $U_{\alpha}n_{(i)}^{\alpha}=0$, a straightforward computation shows that
\begin{equation}
  n^{\alpha}_{(r)}=\left(0,\sqrt{\frac{\Sigma}{r}},0,0\right),\quad
  n^{\alpha}_{(\chi)}=\left(0,0,\frac{\sqrt{1-\chi^2}}{\kappa},0\right),\quad
  n^{\alpha}_{(\varphi)}=\left(0,0,0,\frac{1}{\kappa\sqrt{1-\chi^2}}\right).
\end{equation}
Finally, the non-trivial Einstein field equations are
\begin{eqnarray}
G_{tt}&=&8\pi N^2\widetilde{\rho}+8\pi(1-\chi^2)\omega^2\kappa^2 P_\varphi,\label{Gtt}\\
G_{t\varphi}&=&-8\pi(1-\chi^2)\omega\kappa^2 P_\varphi,\label{Gtphi}\\
G_{\varphi\varphi}&=&8\pi(1-\chi^2)\kappa^2 P_\varphi,\label{Gphiphi}\\
G_{rr}&=&\frac{8\pi r}{\Sigma}P_r,\label{Grr}\\
G_{\chi\chi}&=&\frac{8\pi\kappa^2}{1-\chi^2}P_\chi,\label{Gchichi}\\
G_{r\chi}&=&0.\label{Grchi}
\end{eqnarray}
Equation \eqref{Grchi} can be expressed as
\begin{equation}\label{cond1}
  Q(N,\kappa)\partial_\chi\Sigma+M(N,\kappa,\omega)\Sigma=0    
\end{equation}
with
\begin{eqnarray}
  Q(N,\kappa) &=& \frac{\partial_r(N\kappa)^2}{2},\\
  M(N,\kappa,\omega) &=& 2N\kappa(N\partial_{r\chi}\kappa + \kappa\partial_{r\chi}N) - 2N\partial_r\kappa\partial_\chi(N\kappa) - (1-\chi^2)\kappa^4\partial_r\omega\partial_\chi\omega.
\end{eqnarray}
Notice that $Q\neq 0$ in the limit $J\to 0$. Moreover, equations \eqref{Gtphi} and \eqref{Gphiphi} form an overdetermined linear algebraic system for the single unknown $P_\varphi$. Requiring this system to be consistent leads to the condition $G_{t\varphi} + \omega G_{\varphi\varphi}=0$ which can be rewritten in the form
\begin{equation}\label{cond2}
  S_4(N,\kappa,\Sigma)\partial_{rr}\omega + S_3(N,\kappa,\Sigma)\partial_{\chi\chi}\omega + S_2(N,\kappa,\Sigma)\partial_r\omega + S_1(N,\kappa,\Sigma)\partial_{\chi}\omega = 0
\end{equation}
with
\begin{eqnarray}
S_1(N,\kappa,\Sigma)&=&-r^2\left\{
(1-\chi^2)N\kappa\partial_\chi\Sigma-2\Sigma\left[2(1-\chi^2)N\partial_\chi\kappa-\kappa((1-\chi^2)\partial_\chi N+4\chi N)\right]\right\},\\
S_2(N,\kappa,\Sigma)&=&\kappa^2\Sigma\left\{rN\kappa\partial_r\Sigma+\Sigma\left[8rN\partial_r\kappa-\kappa(N+2r\partial_r N)\right]\right\},\\
S_3(N,\kappa,\Sigma)&=&2r^2(1-\chi^2)N\kappa\Sigma,\\
S_4(N,\kappa,\Sigma)&=&2rN\kappa^2\Sigma^2.
\end{eqnarray}
If we set $N=1$ (vanishing redshift) and choose $\kappa = r$, so that $K=1$, $\Sigma = r - b(r)$ and $\omega(r,\chi)=\omega(r)$, then condition \eqref{cond1} is automatically satisfied, whereas \eqref{cond2} reduces to
\begin{equation}\label{odeomega}
 2r^2(r-b)\ddot{\omega}+\left[r^2(1-\dot{b})+7r(r-b)\right]\dot{\omega}=0, 
\end{equation}
where the dot denotes differentiation with respect to $r$. In the case of the Morris--Thorne wormhole $b=r_0^2/r$, we can introduce the proper radial distance $\ell = \pm\sqrt{r^2-r_0^2}$ by means of which \eqref{odeomega} reduces to the simpler and singularity free form
\begin{equation}\label{reduced}
  \omega^{''}(\ell)+\frac{4\ell}{\ell^2+r_0^2}\omega^{'}(\ell)=0.    
\end{equation}
Notice that the above equation implies $\omega^{''}(0)=0$ at the throat. Here, the prime denotes differentiation with respect to $\ell$. Its general solution is
\begin{equation}
  \omega(\ell)=c_1+c_2\left[\frac{\ell}{2r_0^2(\ell^2+r_0^2)}+\frac{1}{2r_0^3}\arctan{\left(\frac{\ell}{r_0}\right)}\right], 
\end{equation}
with asymptotic expansion
\begin{equation}
  \omega(\ell)=c_1+\frac{\pi c_2}{4r_0^3}-\frac{c_2}{3\ell^3}+\mathcal{O}\left(\frac{1}{\ell^5}\right).    
\end{equation}
Requiring that $\omega\to 2J/\ell^3$ as $\ell\to\infty$ fixes the integration constants to $c_1=3\pi J/(2r_0^3)$ and $c_2=-6J$. Hence, we end up with the solution
\begin{equation}\label{omega1MT}
  \omega(\ell)=J\Omega(\ell),\quad
  \Omega(\ell)=\frac{3\pi}{2r_0^3}-\frac{6}{2r_0^2}\left[\frac{\ell}{\ell^2+r_0^2}+\frac{1}{r_0}\arctan{\left(\frac{\ell}{r_0}\right)}\right].
\end{equation}
A few remarks are in order. First, the choice of integration constants can be imposed on only one asymptotic end. To render both ends asymptotically Minkowski, one must introduce two coordinate charts, $(t,\ell_+,\chi,\varphi_+)$ and $(t,\ell_-,\chi,\varphi_-)$ related by a rigid rotation 
\begin{equation}
  \varphi_{-}=\varphi_+-\frac{3\pi J}{r_0^3}t.
\end{equation}
In what follows, we restrict attention to the upper asymptotic region with $0\leq\ell<\infty$. Finally, differentiating \eqref{omega1MT} gives $\omega^{''}(\ell)=24\ell/(\ell^2+r_0^2)^3$ which satisfies $\omega^{''}_1(0)=0$ as expected. Moreover, at the throat $\omega(0)=3\pi J/(2r_0^3)$. Hence, the final form of the line element associated with the spinning Morris--Thorne wormhole is
\begin{equation}\label{metricell}
ds^2=-\left[1-J^2\Omega^2(\ell)(1-\chi^2)(\ell^2+r_0^2)\right]dt^2+d\ell^2-2J\Omega(\ell)(1-\chi^2)(\ell^2+r_0^2)dtd\varphi+(\ell^2+r_0^2)\left[\frac{d\chi^2}{1-\chi^2}+(1-\chi^2)d\varphi^2\right]
\end{equation}
with $\Omega(\ell)$ given by \eqref{omega1MT}. From a straightforward \textsc{Maple} computation, we obtain closed-form expressions for the Ricci scalar and the Kretschmann invariant, namely 
\begin{eqnarray}\label{Ricci}
R&=&\frac{18J^4(1-\chi^{2})-2r_{0}^{2}(\ell^{2}+r_{0}^{2})}{(\ell^{2}+r_{0}^{2})^{3}},\\
\mathcal{K}&=&\frac{3564J^{8}(1-\chi^{2})^{2}-72J^{4}(\ell^{2}+r_{0}^{2})\left[2\ell^{2}(2\chi^{2}+1)+r_{0}^{2}(7\chi^{2}-1)\right]+12r_{0}^{4}(\ell^{2}+r_{0}^{2})^{2}}{(\ell^{2}+r_{0}^{2})^{6}}.
\end{eqnarray}
Several sanity checks follow immediately. First, the $\theta$-dependence enters only through even powers of $\chi = \cos\theta$, which implies invariance under $\chi \to -\chi$ and therefore confirms the expected reflection symmetry with respect to the equatorial plane. Moreover, the rotational contributions vanish on the symmetry axis $\chi = \pm1$ and are maximal on the equatorial plane $\chi = 0$. Second, the nonrotating limit is smooth and reproduces the expected static Morris--Thorne behaviour, \emph{i.e.} negative scalar curvature supported by exotic matter, and a positive curvature-squared invariant decaying asymptotically as $|\ell|^{-8}$. Third, regularity at the throat is manifest, \emph{i.e.} at $\ell=0$ both scalars remain finite, so the rotating deformation does not introduce curvature singularities at the throat. As an additional consistency check, we evaluated the Komar angular-momentum charge associated with the axial Killing vector field. The resulting surface integral reproduces exactly $J$, confirming that this parameter can indeed be interpreted as the total angular momentum of the spacetime. Finally, the density and pressures are computed to be
\begin{eqnarray}
\widetilde{\rho}(\ell,\chi)&=&-\frac{r_0^2(\ell^2+r_0^2)+9J^4(1-\chi^2)}{8\pi(\ell^2+r_0^2)^3},\label{rho}\quad
P_R(\ell,\chi)=-\frac{r_0^2(\ell^2+r_0^2)-9J^4(1-\chi^2)}{8\pi(\ell^2+r_0^2)^3},\\
P_t(\ell,\chi)&=&\frac{r_0^2(\ell^2+r_0^2)-9J^4(1-\chi^2)}{8\pi(\ell^2+r_0^2)^3},\quad
P_\varphi(\ell,\chi)=\frac{r_0^2(\ell^2+r_0^2)-27J^4(1-\chi^2)}{8\pi(\ell^2+r_0^2)^3}.\label{Pphi}
\end{eqnarray}
The matter content implied by the rotating configuration inherits the expected reflection symmetry with respect to the equatorial plane. Hence, the stress–energy components are invariant under $\chi\to-\chi$ and reduce to their static Morris–Thorne form on the symmetry axis. In the nonrotating limit, one recovers the standard relations $P_R=\widetilde{\rho}$ and $P_t=P_\varphi=-\widetilde{\rho}$, while for $J\neq0$ the identity $P_t(\ell,\chi)=-P_R(\ell,\chi)$ holds exactly, signaling a persistent radial–polar anti-symmetry and an additional azimuthal anisotropy through $P_\varphi$. Moreover, the NEC conditions are clearly violated. $\widetilde{\rho}(\ell,\chi)<0$ everywhere, and the radial null combination is strictly negative, namely
\begin{equation}\label{NEC1}
  \widetilde{\rho}+P_R=-\frac{r_0^2}{4\pi(\ell^2+r_0^2)^2}<0,
\end{equation}
independently of $J$ and $\chi$. Away from the axis, rotation further strengthens the violation in the angular directions since 
\begin{equation} \label{41c}
\widetilde{\rho} + P_t = -\frac{9J^4(1-\chi^2)}{4\pi(\ell^2+r_0^2)^3} \leq 0,\quad
\widetilde{\rho} + P_\varphi = -\frac{9J^4(1-\chi^2)}{2\pi(\ell^2+r_0^2)^3} \leq 0
\end{equation}
with equality only for $\chi = \pm1$ or $J = 0$. We also notice that for the anisotropic fluid considered here, the weak and strong energy conditions (WEC, SEC) reduce to simple algebraic inequalities in the orthonormal tetrad comoving with the fluid (see, for instance, \cite{Wald1986})
\begin{eqnarray}
\text{WEC} &:&\quad \widetilde{\rho} \geq 0,\quad \widetilde{\rho} + P_R \geq 0, \quad \widetilde{\rho} + P_t \geq 0,\quad \widetilde{\rho} + P_\varphi \geq 0,\\
\text{SEC} &:&\quad \widetilde{\rho}+P_R+P_t+P_\varphi\geq 0,\quad \widetilde{\rho} + P_R \geq 0,\quad \widetilde{\rho} + P_t \geq 0,\quad \widetilde{\rho} + P_\varphi \geq 0.
\end{eqnarray}
Using \eqref{rho}-\eqref{Pphi} yields
\begin{equation}\label{41e}
\widetilde{\rho} + P_R + P_t + P_\varphi = -\frac{9J^4(1-\chi^2)}{2\pi(\ell^2+r_0^2)^3} \leq 0. 
\end{equation}
Notice that the angular null combinations \eqref{41c} and the SEC sum \eqref{41e} become strictly negative away from the symmetry axis whenever $J \neq 0$, while equality is attained only on the axis or in the static limit $J = 0$. Consequently, the NEC is violated in all principal directions throughout the spacetime, including along the axis, in agreement with the generic requirement of exotic matter for traversable wormholes \cite{Morris1988AJP}. Since $\widetilde{\rho}<0$ everywhere, the WEC and therefore also the dominant energy condition are violated globally. The SEC is also violated wherever $J\neq 0$ and $|\chi| < 1$, as indicated by \eqref{41e}. Finally, the trace relation provides a further sanity check. Using $T = -\widetilde{\rho}+P_R+P_t+P_\varphi=-\widetilde{\rho} + P_\varphi$ one obtains $R = -8\pi T$, which reproduces the Ricci scalar computed in \eqref{Ricci}. Asymptotically, the $J$-dependent contributions fall off faster, \emph{i.e.} like $|\ell|^{-6}$, than the static terms decaying like $|\ell|^{-4}$. As expected, rotation mainly affects the near-throat off-axis region.

\section{Main results}

\subsection{Shadows of the rotating wormhole}

\begin{figure}[H]
  \begin{subfigure}[b]{0.23\textwidth}
    \centering
    \includegraphics[width=\textwidth]{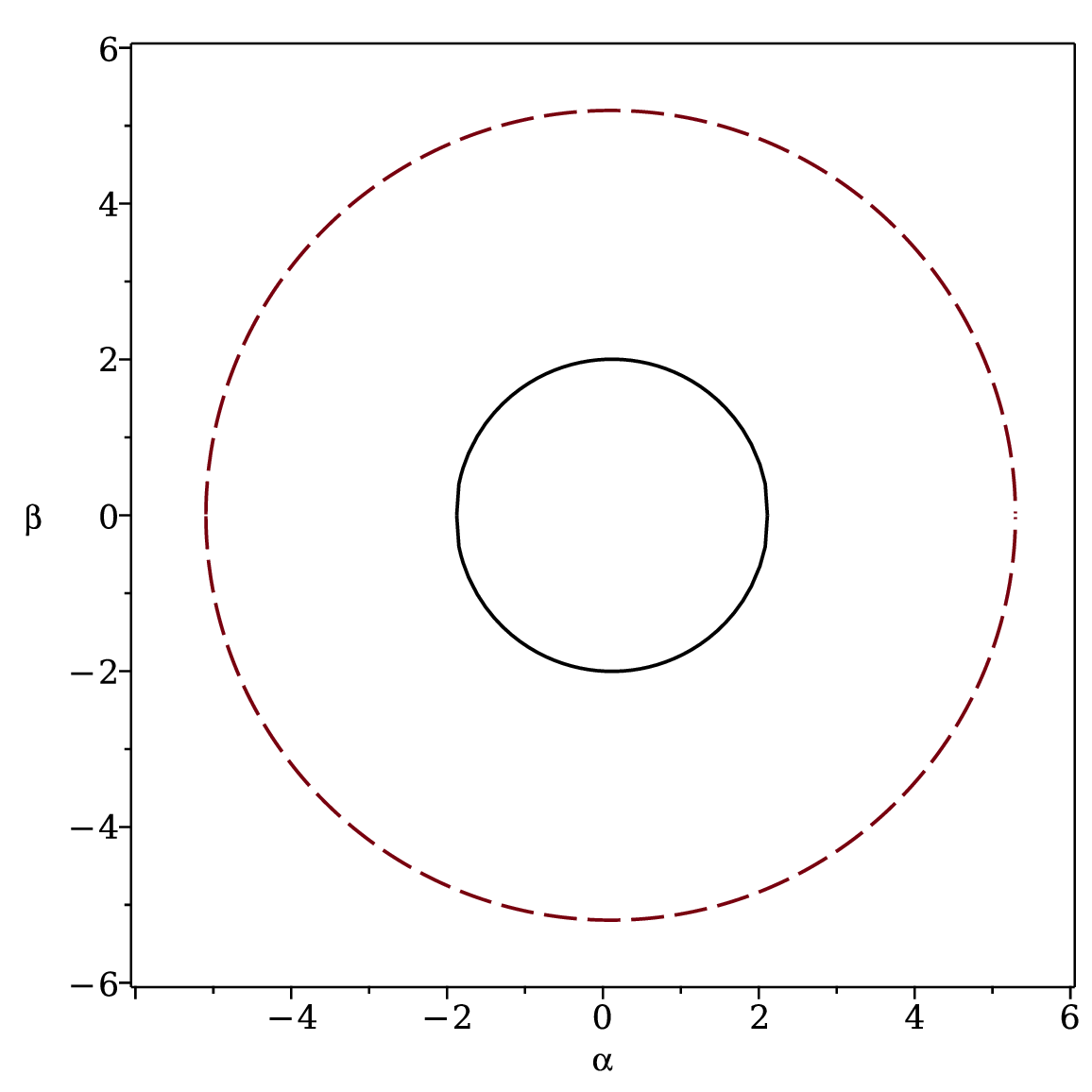}
    \caption{$J=0.05, \theta_0 = 90^\circ$}
  \end{subfigure}
  \hfill
  \begin{subfigure}[b]{0.23\textwidth}
    \centering
    \includegraphics[width=\textwidth]{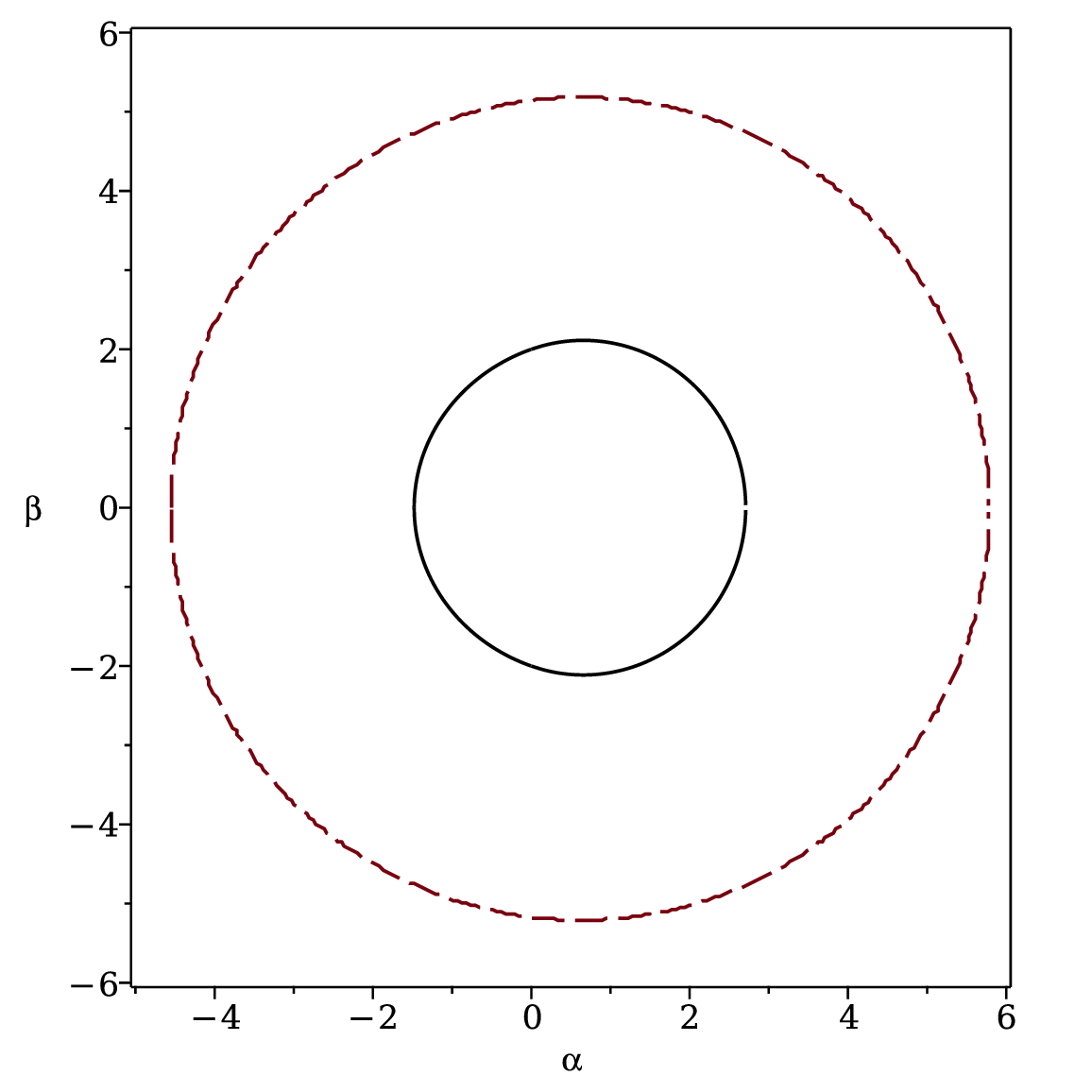}
    \caption{$J=0.3, \theta_0 = 90^\circ$}
  \end{subfigure}
  \hfill
  \begin{subfigure}[b]{0.23\textwidth}
    \centering
    \includegraphics[width=\textwidth]{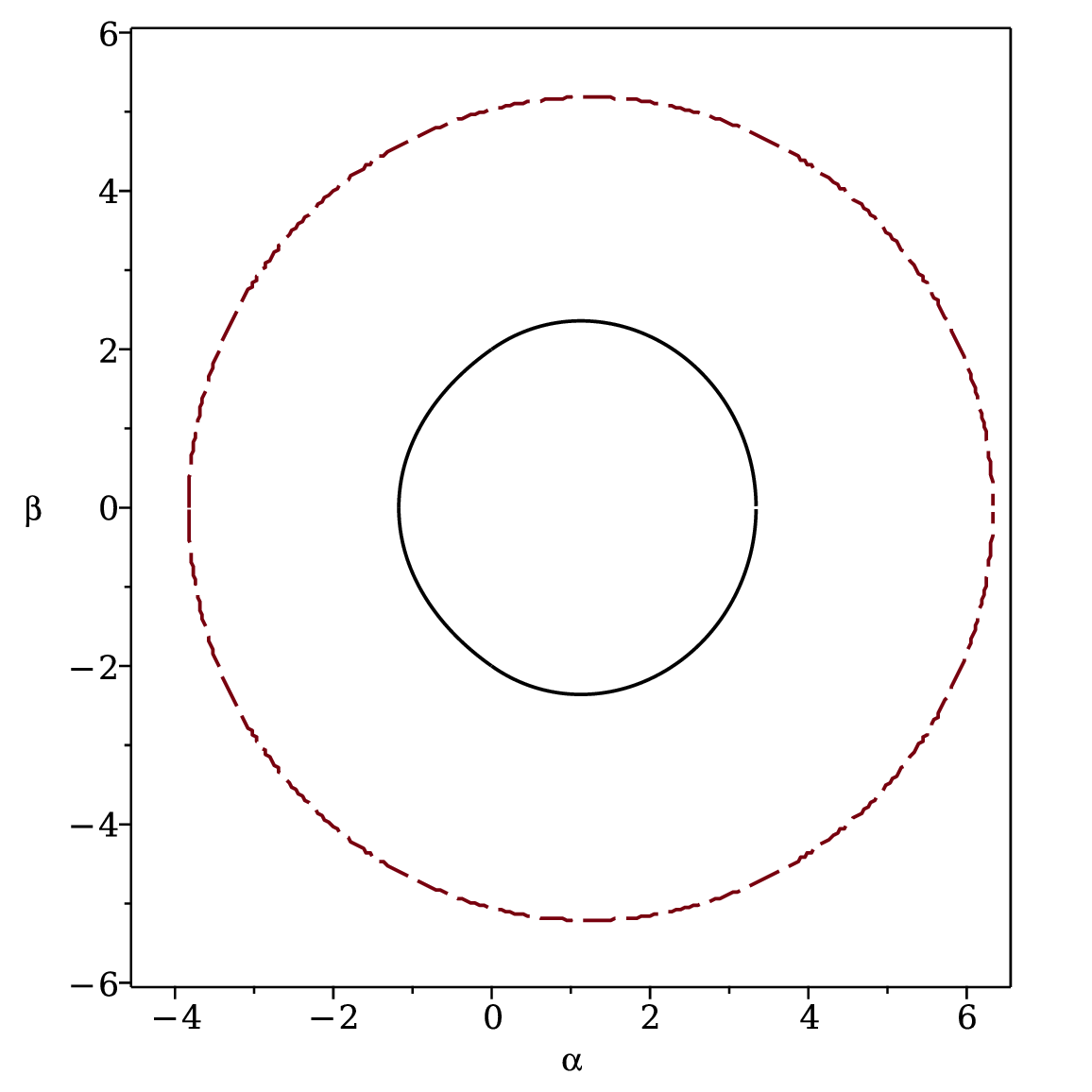}
    \caption{$J=0.6, \theta_0 = 90^\circ$}
  \end{subfigure}
  \hfill
  \begin{subfigure}[b]{0.23\textwidth}
    \centering
    \includegraphics[width=\textwidth]{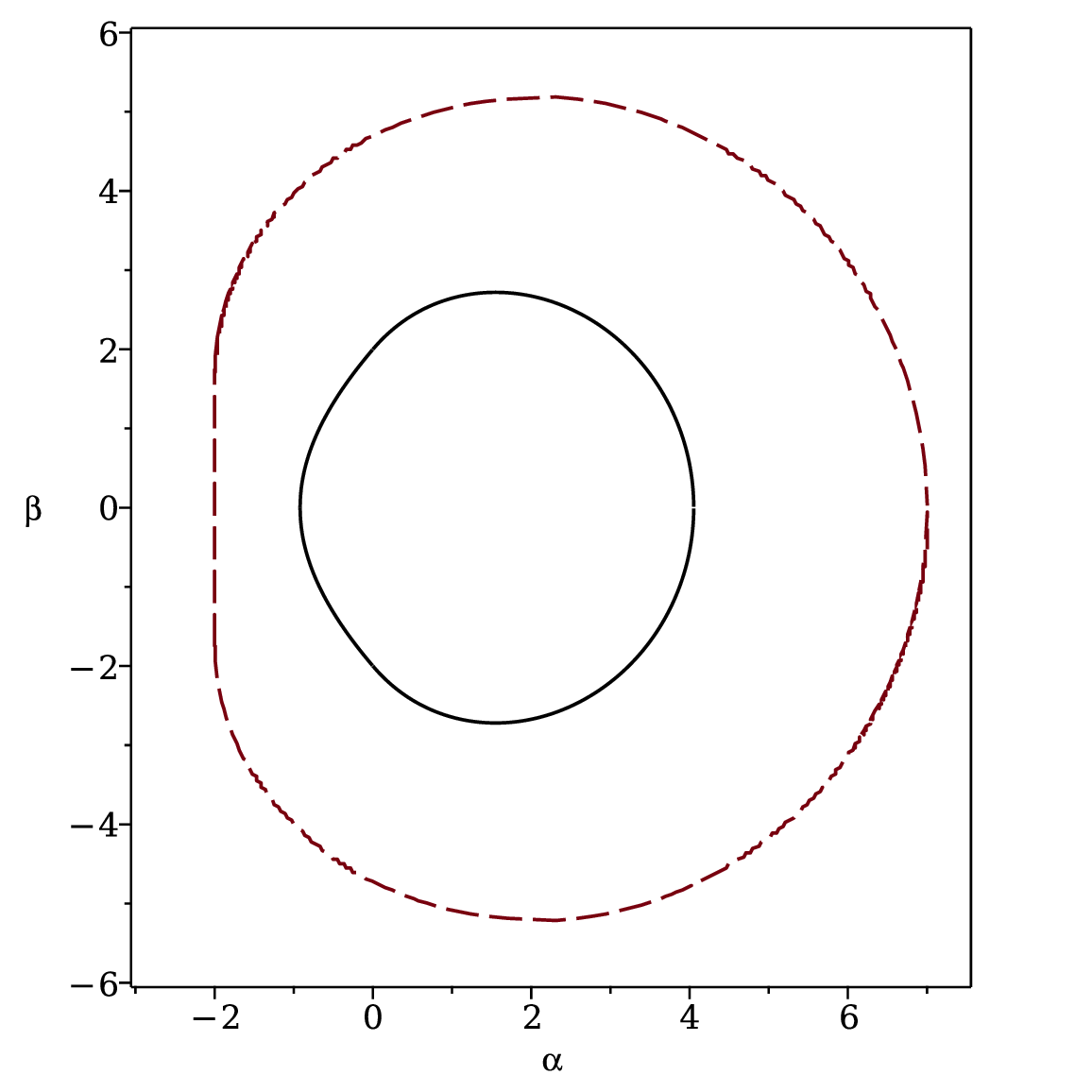}
    \caption{$J=1, \theta_0 = 90^\circ$}
  \end{subfigure}

  \vspace{1em}

  \begin{subfigure}[b]{0.23\textwidth}
    \centering
    \includegraphics[width=\textwidth]{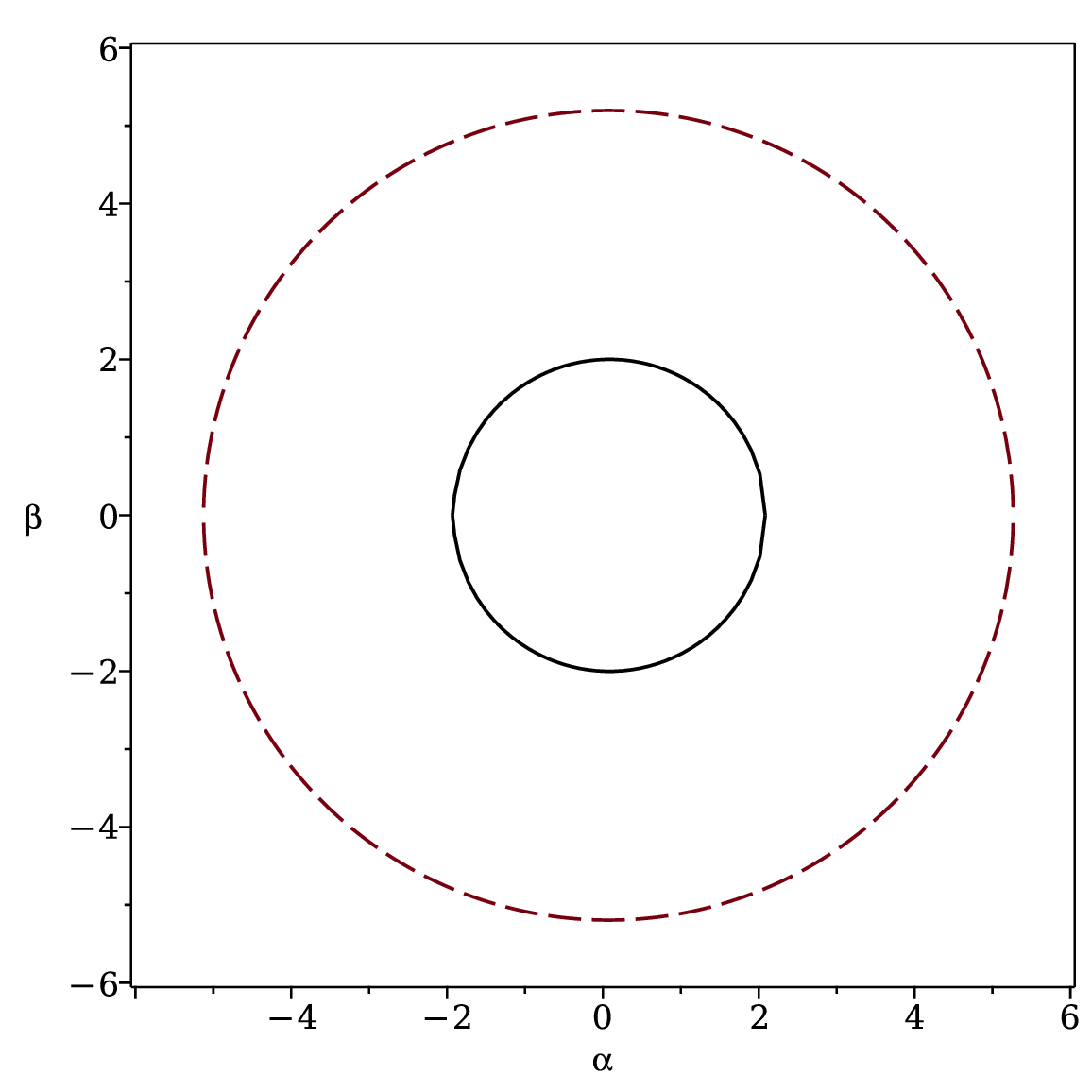}
    \caption{$J=0.05, \theta_0 = 45^\circ$}
  \end{subfigure}
  \hfill
  \begin{subfigure}[b]{0.23\textwidth}
    \centering
    \includegraphics[width=\textwidth]{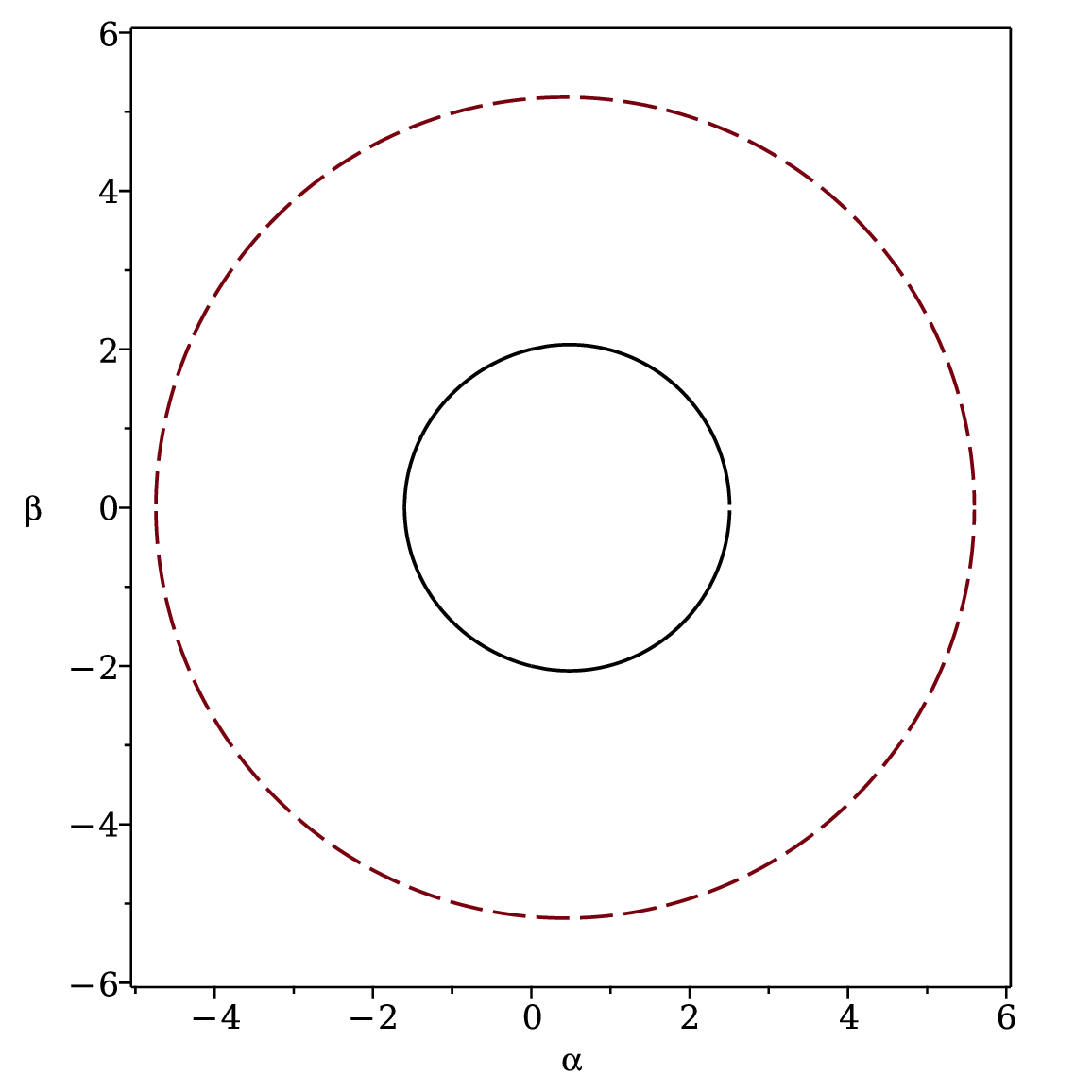}
    \caption{$J=0.3, \theta_0 = 45^\circ$}
  \end{subfigure}
  \hfill
  \begin{subfigure}[b]{0.23\textwidth}
    \centering
    \includegraphics[width=\textwidth]{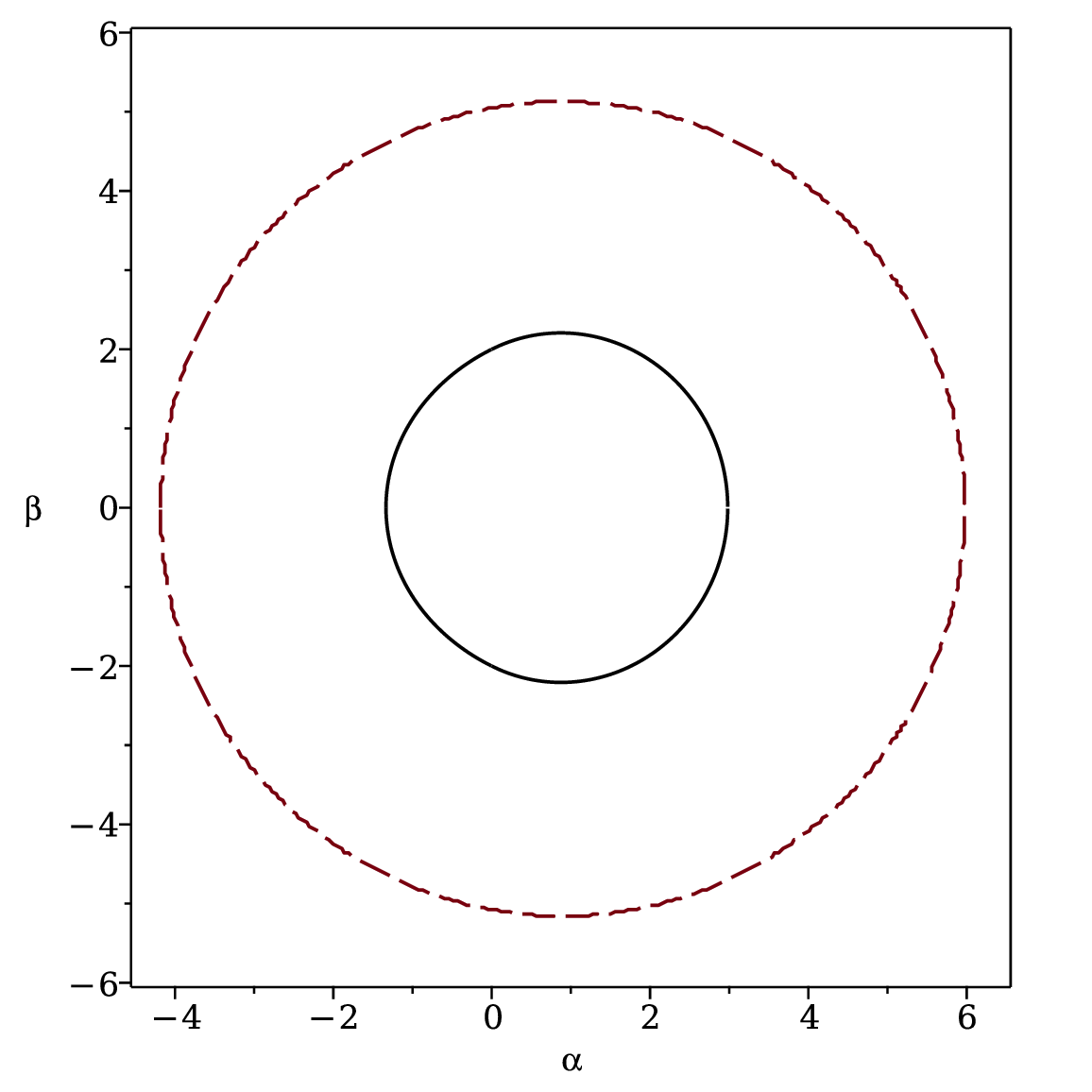}
    \caption{$J=0.6, \theta_0 = 45^\circ$}
  \end{subfigure}
  \hfill
  \begin{subfigure}[b]{0.23\textwidth}
    \centering
    \includegraphics[width=\textwidth]{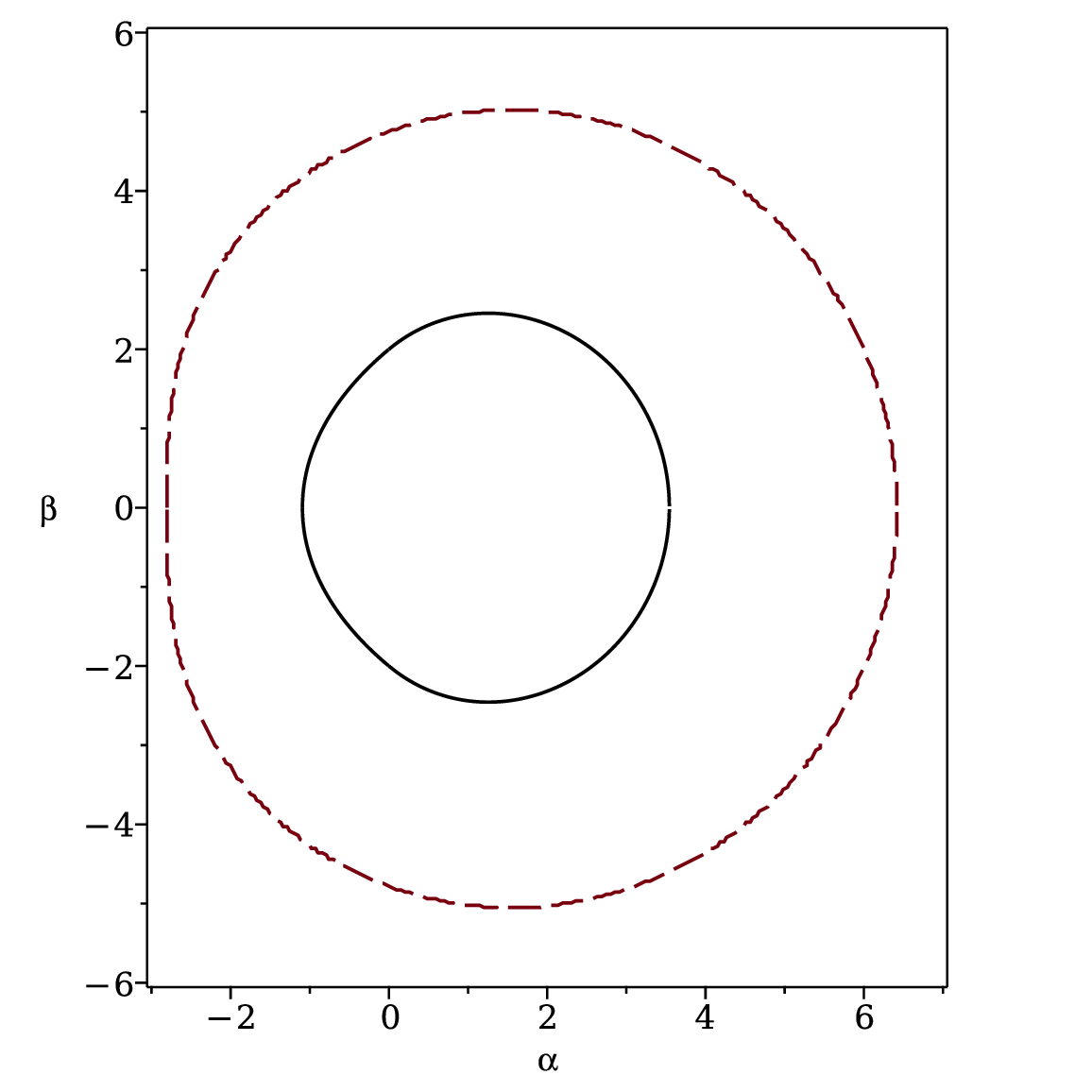}
    \caption{$J=1, \theta_0 = 45^\circ$}
  \end{subfigure}

  \caption{The shadow of a rotating Morris--Thorne wormhole (solid black line) with shape function $b(r) = r_0^2/r$, and the Kerr black hole (dashed red line) for different spin values and different inclination angles. Here, the mass of the Kerr solution is set to 1 and $r_0 = 2$ for the Morris--Thorne wormhole. The coordinates are in units of mass.}
  \label{fig:shadowMT}
\end{figure}

The shadow of a rotating traversable wormhole of Teo type was first analysed in detail in \cite{Petya2013PRD}, which obtained analytic expressions for the boundary of the shadow in terms of the conserved impact parameters of null geodesics. A later reanalysis in \cite{Shaikh2018PRD} pointed out that the original treatment neglected the possible contribution of unstable photon orbits at the throat and derived the correct shadow boundary by including these orbits explicitly. Related aspects of geodesic motion and strong‑field observables for this class of geometries have been further explored, for instance, in the context of quasiperiodic oscillations around rotating wormholes in \cite{Deligianni2021PRD}. Following the formulation of \cite{Petya2013PRD, Shaikh2018PRD}, we consider the class of Teo‑type wormhole metrics for which the functions $N(r)$, $K(r)$ and $\omega(r)$ depend only on the areal radius $r$. In this setting, the null geodesic equations admit two conserved impact parameters, $\xi = L/E$, and $\eta = Q/E^2$, where $E$, $L$ and $Q$ are the energy, axial angular momentum and Carter constant of the photon orbit. For an observer located at spatial infinity at inclination angle $\theta_0$ with respect to the rotation axis, the apparent position of a photon on the observer’s sky is described by the celestial coordinates $(\alpha,\beta)$, which are given by \cite{Petya2013PRD}
\begin{equation}
  \alpha(r_{ph}) = -\frac{\xi(r_{ph})}{\sin\theta_0}, 
  \qquad
  \beta(r_{ph}) = \pm\sqrt{\eta(r_{ph}) 
  - \frac{\xi^2(r_{ph})}{\sin^2\theta_0}},
  \label{eq:shadow-outside}
\end{equation}
for $r_{ph} \geq r_0$. Here $r_{ph}$ denotes the radius of an unstable spherical photon orbit, and $r_0$ is the throat radius. Equation~\eqref{eq:shadow-outside} describes the part of the shadow boundary generated by unstable spherical photon orbits lying strictly outside the throat, $r_{ph} > r_0$. The critical values of $(\xi, \eta)$ associated with these unstable spherical orbits follow from the conditions $R(r_{ph}) = 0$ and $R'(r_{ph}) = 0$ on the radial potential $R(r)$, and can be written in parametric form as \cite{Petya2013PRD}
\begin{equation}
  \eta(r) 
  = \frac{r^{2}K^{2}(r)}{N^{2}(r)}\bigl[1-\omega(r)\,\xi(r)\bigr]^{2},
  \qquad
  \xi(r) 
  = \frac{\Sigma(r)}{\Sigma(r)\,\omega(r)-\omega'(r)},
  \label{eq:xi-eta}
\end{equation}
with
\begin{equation}
  \Sigma(r) = \frac{1}{2}\,\frac{d}{dr}
  \ln\!\left(\frac{N^{2}(r)}{r^{2}K^{2}(r)}\right).
  \label{eq:Sigma-def}
\end{equation}
In our notation $N(r)$ is the lapse function, $b(r)$ not appearing explicitly in the above formulae is the shape function, $K(r)$ encodes the deviation of the $r=$const two‑spheres from spherical geometry, and $\omega(r)$ is the angular velocity of inertial frames in the asymptotic region. As emphasised in \cite{Shaikh2018PRD}, the throat can itself carry unstable photon orbits and hence contributes nontrivially to the shadow boundary. For circular photon orbits located at the throat $r = r_0$, the radial conditions reduce to $R(r_0) = 0$, and combining this with the definition of the celestial coordinates leads to a quadratic relation between $\alpha$ and $\beta$, namely \cite{Shaikh2018PRD}
\begin{equation}
  \bigl[N_0^{2} - \omega_0^{2} r_0^{2} K_0^{2}\sin^{2}\theta_0\bigr]\alpha^{2}
  - 2 \omega_0 r_0^{2} K_0^{2}\sin\theta_0\,\alpha
  - r_0^{2}K_0^{2} + N_0^{2}\beta^{2} = 0,
  \label{eq:throat-quadratic}
\end{equation}
where $N_0=N(r_0)$, $K_0=K(r_0)$ and $\omega_0=\omega(r_0)$. Solving \eqref{eq:throat-quadratic} for $\beta$ yields the additional segment of the shadow boundary associated with throat photon orbits,
\begin{equation}
  \beta(\alpha) 
  = \pm\frac{1}{N_0}\sqrt{
  r_0^{2} K_0^{2}\bigl(1 + \omega_0 \alpha \sin\theta_0\bigr)^{2}
  - N_0^{2}\alpha^{2}}.
  \label{eq:shadow-throat}
\end{equation}
The full shadow is then obtained as the envelope of the curves defined by \eqref{eq:shadow-outside}, \eqref{eq:xi-eta}--\eqref{eq:Sigma-def}, and \eqref{eq:shadow-throat}, retaining only those segments that correspond to unstable photon orbits and lie on the boundary between capture and escape trajectories~\cite{Shaikh2018PRD}. Figure~\ref{fig:shadowMT} compares the resulting shadows for the rotating Morris--Thorne wormhole with shape function $b(r)=r_0^{2}/r$ to the Kerr black hole with the same mass for various spins and inclinations. In all the examples shown, the wormhole shadow is smaller than the corresponding Kerr shadow. This demonstrates explicitly that the larger‑than‑Kerr shadows reported for some Teo‑type wormholes in \cite{Petya2013PRD} are not generic but rather depend on the specific choice of metric functions. Moreover, \eqref{eq:xi-eta} shows that, while the shape function $b(r)$ does not enter explicitly in the expression for the critical impact parameters, it affects the shadow indirectly through its influence on $N(r)$, $K(r)$ and, most importantly, on the rotation function $\omega(r)$, once the full set of Einstein equations and matter constraints are imposed. In our exact spinning Morris--Thorne configuration, the rotation profile $\omega(r)$ is determined by the throat geometry and thus inherits a nontrivial dependence on $r_0$. As a consequence, the shadow size and shape depend on $r_0$ as well. This implies that, at least for this class of analytic rotating wormholes, the shadow can in principle probe not only the angular momentum $J$ but also the properties of the throat encoded in the shape function.

\begin{figure}[H]
\centering
    \includegraphics[width=0.45\textwidth]{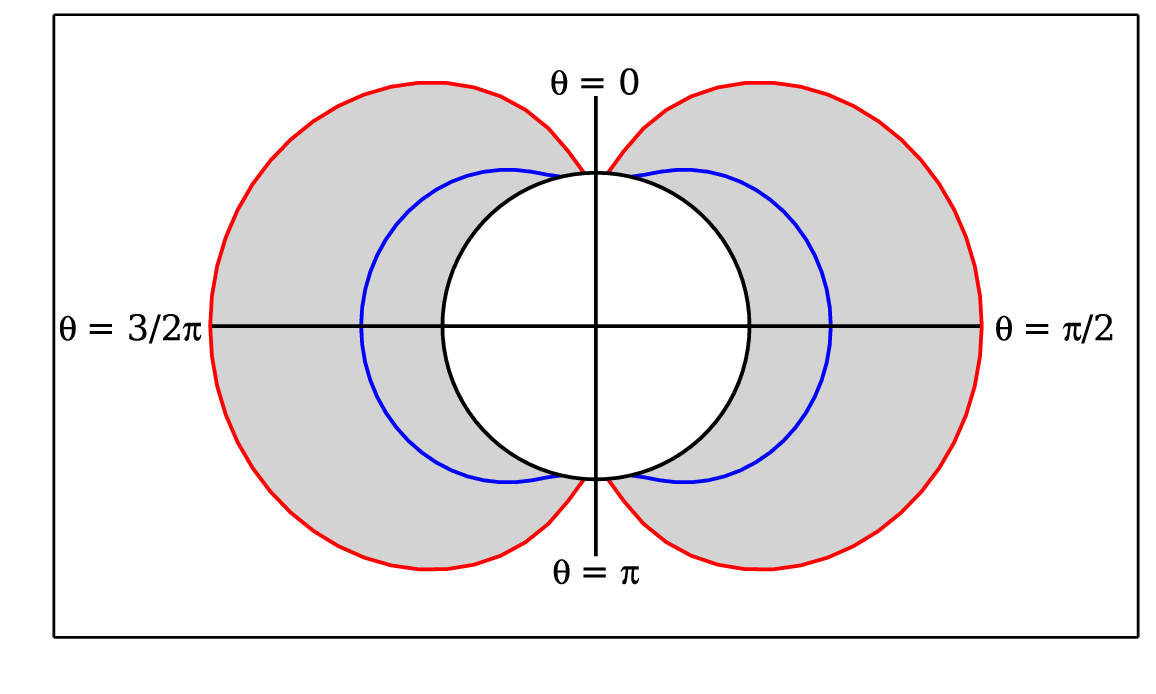}
    \caption{\label{ERGO} Cross section of the wormhole throat for a rotating Morris--Thorne wormhole with shape function $b(r) = r_0^2/r$. The solid blue and red curves correspond to the boundary of the ergosphere corresponding to the rescaled spin parameters $j=1$ and $j=3$, respectively, while the solid black line represents the throat of the wormhole.}
\end{figure}

\subsection{Ergoregion}

We show that the spacetime admits a critical spin 
\begin{equation}\label{Jc}
  |J_c|=\frac{2r_0^2}{3\pi},
\end{equation}
which separates qualitatively different regimes. For $|J| < |J_c|$, the timelike Killing vector $\partial_t$ remains timelike everywhere and no ergoregion is present. At the critical value $|J| = |J_c|$, the ergosurface collapses to the throat, and hence, it cannot form an open ergoregion. Only for $|J| > |J_c|$ does a genuine ergoregion develop around the throat, bounded by a smooth ergosurface. Let us recall that for a stationary, axisymmetric spacetime with time‑translation Killing vector $\xi^\alpha=\partial_t$, the ergoregion $E$ is the set where $\xi^\alpha$ becomes spacelike, i.e. $E=\{p\in \mathcal{M}~|~ g_{tt}(p)>0\}$, and its boundary, that is the ergosurface,  is where $g_{tt}=0$. For our spinning Morris--Thorne wormhole, the metric in $(t,\ell,\chi,\phi)$ coordinates is given by \eqref{metricell} with the rotation profile $\omega(\ell)=J\Omega(\ell)$ specified by \eqref{omega1MT}. Let us recall that $g_{tt}(\ell,\chi)
= -\left[1 - J^2\Omega(\ell)^2(1-\chi^2)(\ell^2+r_0^2)\right]$ and define the non‑negative function $H(\ell,\chi):=\Omega(\ell)^2(1-\chi^2)(\ell^2+r_0^2)$, so that $g_{tt} = -(1 - J^2 H)$. Then, $g_{tt}<0$ for $J^2H<1$, $g_{tt}=0$ whenever $J^2H=1$, and $g_{tt}>0$ for $J^2H>1$. Thus, an ergoregion exists if and only if the supremum of $H$  satisfies
\begin{equation}
  J^2\sup_{\ell\in\mathbb{R},~\chi\in[-1,1]}\{H(\ell,\chi)\} > 1.
\end{equation}
In order to evaluate the supremum, we start observing that from the reduced field equation \eqref{reduced} and its solution \eqref{omega1MT}, we have 
\begin{equation}\label{omegap}
  \Omega^{'}(\ell)= -\frac{6}{(\ell^2+r_0^2)^2}.
\end{equation}
Thus $\Omega^{'}(\ell)<0$ for all $\ell$, so $\Omega$ is strictly decreasing on $[0,\infty)$. From \eqref{omega1MT}, at the throat $\Omega(0) = 3\pi/(2r_0^3)$, and moreover, $\Omega(\ell)\longrightarrow 0$ as $|\ell|\to\infty$. It is not difficult to check that $\Omega(\ell)>0$ for all finite $\ell$. So on each asymptotic end $\ell\geq 0$ or $\ell\leq 0$, $\Omega(\ell)$ is a positive, strictly decreasing function from $\Omega(0)=3\pi/(2r_0^3)$ down to $0$. On the other hand, the factor $1-\chi^2$ is maximised at the equatorial plane $\chi=0$. So for each $\ell$, $\max_{\chi} H(\ell,\chi) = H(\ell,0) = \Omega(\ell)^2(\ell^2 + r_0^2)$. Thus, the supremum of $H$ is obtained at some point $(\ell_*,0)$, and we only need to study the 1‑variable function $\mathcal{B}(\ell) = \Omega(\ell)^2(\ell^2 + r_0^2)$ for $\ell\in\mathbb{R}$. By symmetry of the wormhole, $\mathcal{B}$ is even in $\ell$, so we can restrict to $\ell \geq 0$. To verify the monotonicity of $\mathcal{B}$, we observe that
\begin{equation}
  \mathcal{B}^{'}(\ell)=2\Omega(\ell)F(\ell),\quad 
  F(\ell)=\ell\Omega(\ell)-\frac{6}{\ell^2+r_0^2}
\end{equation}
where we used \eqref{omegap}. Then $B^{'}(\ell)$ has the same sign as $F(\ell)$, since $\Omega(\ell)>0$. At the throat, $F(0)=-6/r_0^2<0$. Moreover, for large $\ell$, we find $F(\ell)\sim-4/\ell^2<0$. A straightforward computation shows that
\begin{equation}
  F^{'}(\ell)=\Omega(\ell)+\frac{6\ell}{(\ell^2+r_0^2)^2}>0
\end{equation}
for every $\ell\geq 0$, because the first term is always positive. Hence, $F(\ell)$ is strictly increasing on $[0,\infty)$, yet $F(0)<0$ and $F(\ell\to 0^-$ as $\ell\to\infty$. Hence $F(\ell)<0$ for all $\ell>0$, and therefore $\mathcal{B}^{'}(\ell)<0$ for all  $\ell\leq 0$. We conclude that $\mathcal{B}(\ell)$ is strictly decreasing on $[0,\infty)$, with maximum at the throat given by $B_{\max}=9\pi^2/(4r_0^4)$. Since $\mathcal{B}$ is even, this is also the maximum over all $\ell\in\mathbb{R}$. Finally, the global supremum of $H(\ell,\chi)$ is  $9\pi^2/(4r_0^4)$ and it is attained at the equatorial throat point $(\ell=0, \chi=0)$. This leads to the critical spin value \eqref{Jc} and an ergoregion will exist provided that $|J| > J_c$.  If you define a dimensionless spin parameter $j = J/r_0^2$, the critical value is $j_c\approx 0.212$. For $|J| > |J_c|$, the ergosurface is given implicitly by $g_{tt} = 0$, 
$J^2\Omega(\ell)^2(1 - \chi^2)(\ell^2 + r_0^2) = 1$. Because $\mathcal{B(\ell)}$ is maximal at the throat and strictly decreasing with $|\ell|$, the ergosphere equation describes, on each side of the throat, a closed surface around the throat. For fixed $J > |J_c|$, there is a finite interval $|\ell|<\ell_\text{max}(J)$ where $J^2\mathcal{B}(\ell)\geq 1$, inside this, $g_{tt}>0$, and the ergosurface intersects the equatorial plane at $|\ell|=\ell_\text{max}(J)$. At the throat $(\ell=0)$, the intersection is at latitudes $\chi=\pm\chi_0(J)$ defined by $1-\chi_0^2 = 1/(J^2\mathcal{B}_{\max})$. This matches the bubble around the throat we displayed in Fig.~\ref{ERGO} for several values of the rescaled spin parameter. Finally, there is no ergosphere for small rotation. More precisely, for all $|J|<|J_c|$, the timelike Killing vector $\partial_t$ remains timelike everywhere, so the ergoregion is empty.

\subsection{Causal Structure}

As discussed in the previous subsection, the metric component $g_{tt}$ admits a global maximum at the equatorial throat, leading to a critical spin $|J_c| = 2r_0^2/(3\pi)$ such that $g_{tt}<0$ everywhere and no ergoregion is present for $|J| < |J_c|$, whereas an ergoregion develops around the throat when $|J| > |J_c|$. In rotating spacetimes, the presence of an ergoregion often raises questions about the possible formation of closed timelike curves (CTCs). We now show that, in the present geometry, no CTCs exist for any value of $J$. Since CTCs are closed curves whose tangent vector is everywhere timelike, it is natural to begin by examining curves generated by the Killing symmetries of the spacetime. Consider first the azimuthal Killing vector $\partial_\varphi$. From the line element \eqref{metricell} we have $g_{\varphi\varphi} = (\ell^2 + r_0^2)(1-\chi^2)$, which is strictly positive for $|\chi|<1$ and vanishes only on the symmetry axis, where the $\varphi$-circles shrink to a point, and the usual coordinate degeneracy occurs. Thus, the closed orbits of $\partial_\varphi$ are everywhere spacelike, and purely azimuthal CTCs are excluded. One may next consider helical Killing vectors of the form $k^\alpha = a\partial_t + b\partial_\varphi$. Along an orbit of $k^\alpha$ at fixed $(\ell,\chi)$ we have $t(\lambda) = a\lambda + t_0$ and $\varphi(\lambda) = b\lambda + \varphi_0$. For the orbit to close, one needs $t(\lambda_0)=t_0$ and $\varphi(\lambda_0)=\varphi_0 + 2\pi n$ for some $\lambda_0\neq 0$ and integer $n$. The first condition forces $a\lambda_0=0$, and hence $a=0$, since we do not periodically identify the time coordinate. Therefore, the only closed Killing orbits are the pure $\varphi$-circles already shown to be spacelike. Although certain helical combinations $a\partial_t + b\partial_\varphi$ can become timelike inside the ergoregion due to frame dragging, their integral curves are not closed unless $t$ is compactified. Since $t$ is globally defined and noncompact, no CTCs arise from Killing flows alone.

The absence of Killing-generated CTCs does not by itself guarantee global causal regularity, because closed timelike curves could in principle follow more general, non-symmetric paths. To rule these out, we now establish the existence of a global temporal function. Consider the scalar function $t: \mathcal{M} \rightarrow \mathbb{R}$, defined globally on our spacetime $\mathcal{M}$. Its gradient is $\nabla_\alpha t = (dt)_\alpha$, and its norm is $g^{\alpha\beta}(\nabla_\alpha t)(\nabla_\beta t) = g^{tt}$. For the spinning Morris–Thorne metric \eqref{metricell}, it is convenient to rewrite the $(t,\varphi)$ sector in the form
\begin{equation}
  ds^2_{(t,\varphi)} = -dt^2 + A(\ell,\chi)\left(d\varphi - \omega(\ell) dt\right)^2,
\end{equation}
with $A(\ell,\chi) = (\ell^2 + r_0^2)(1-\chi^2)>0$ and $\omega(\ell)=J\Omega(\ell)$. This decomposition makes it manifest that the spacetime is of the ADM type with unit lapse $(N=1)$. The full line element can be written as
\begin{equation}
  ds^2 = -dt^2 + h_{ij}(dx^i + \beta^i dt)(dx^j + \beta^j dt),
\end{equation}
where $x^i = (\ell, \chi, \varphi)$, the spatial metric $h_{ij}$ is positive definite, and the shift has only an azimuthal component proportional to $\omega(\ell)$. In any such $3+1$ decomposition one has the standard identity $g^{tt} =-1/N^2$, so in our case $g^{tt} = -1$ everywhere. Equivalently, one can verify directly from \eqref{metricell} that
\begin{equation}
  g^{tt} = \frac{g_{\varphi\varphi}}{g_{\varphi\varphi}g_{tt}-g_{t\varphi}^2} = -1.
\end{equation}
In either approach, the conclusion is that $g^{\alpha\beta}(\nabla_\alpha t)(\nabla_\beta t) = -1 < 0$ everywhere on the manifold. Thus $\nabla_\alpha t$ is a globally timelike covector field. By choosing the time orientation so that $-\nabla_\alpha t$ is future-directed, $t$ becomes a temporal function in the sense of Lorentzian causality theory, \emph{i.e.} it increases strictly along every future-directed causal curve, and its level sets $t = \text{const}$ are spacelike, acausal hypersurfaces providing a global foliation of the spacetime. By standard results (see, \emph{e.g.}, the discussion of temporal functions and stable causality in \cite{Wald1986}), the existence of such a function implies that the spacetime is stably causal and, in particular, contains no closed timelike curves. We emphasise that this argument does not rely on the sign of $g_{tt}$ and therefore applies equally well in the presence or absence of an ergoregion. Even when $|J| > |J_c|$ and $g_{tt} > 0$ in a neighborhood of the throat, the contravariant component $g^{tt} = -1$ remains negative everywhere, so the causal character of $\nabla_\alpha t$ is unaffected. The ergoregion modifies the behaviour of the timelike Killing vector $\partial_t$ but does not compromise the global causal structure.

Finally, the above reasoning extends to the general Teo ansatz \eqref{rotmet}. Rewriting \eqref{rotmet} as
\begin{equation}
ds^2 = -N(r,\chi)^2 dt^2 + \frac{r}{\Sigma(r,\chi)}dr^2 + \kappa (r,\chi)^2\left[\frac{d\chi^2}{1-\chi^2} + (1-\chi^2)\left(d\varphi-\omega(r,\chi) dt\right)^2\right],
\end{equation}
one again identifies an ADM decomposition with lapse $N(r,\chi)$ and a purely azimuthal shift. Therefore $g^{tt} = -1/N(r,\chi)^2$, which is strictly negative wherever $N(r,\chi)$ is real and nonvanishing. Thus, any Teo-type rotating wormhole with regular, positive lapse is stably causal and cannot develop CTCs, unless one introduces nonstandard global identifications of the time coordinate by hand.

\subsection{Geroch--Hansen multipole structure and comparison with Kerr}

We compute the Geroch--Hansen (GH) multipole moments \cite{Geroch1970JMP, Geroch1970JMPa, Hansen1974JMP} to give a coordinate invariant characterization of the far-field generated by the spinning MT wormhole, to quantify precisely its departure from Kerr revealing a massless, purely spinning source with vanishing mass monopole and quadrupole but non-trivial higher multipoles, and to provide the appropriate input for further studies of orbital dynamics, lensing and gravitational wave emission in this geometry.

For large $\ell$ we can extract the following asymptotic expansions from \eqref{metricell}
\begin{eqnarray}
g_{tt}&=&-1+\frac{4J^2(1-\chi^2)}{\ell^4}+\mathcal{O}\left(\frac{1}{\ell^6}\right),\quad
\omega(\ell)=\frac{2J}{\ell^3}-\frac{12Jr_0^2}{5\ell^5}+\mathcal{O}\left(\frac{1}{\ell^7}\right),\\
g_{t\varphi}&=&-\frac{2J(1-\chi^2)}{\ell}+\mathcal{O}\left(\frac{1}{\ell^3}\right),\quad
g_{\varphi\varphi}=\ell^2(1-\chi^2)+\mathcal{O}(1).
\end{eqnarray}
Since we can identify $\ell$ with the radial coordinate $r$ asymptotically away from the throat and $\chi=\cos{\vartheta}$, we find
\begin{equation}
g_{tt}=-1+\mathcal{O}\left(\frac{1}{r^4}\right),\quad
g_{t\varphi}=-\frac{2J\sin^2{\vartheta}}{r}+\mathcal{O}\left(\frac{1}{r^3}\right).
\end{equation}
If we compare them with the standard asymptotically Cartesian and mass-centred (ACMC) expansion \cite{Thorne1980RMP} holding for any stationary asymptotically flat spacetime, namely
\begin{equation}
  g_{tt}=-1+\frac{2M_0}{r}+\mathcal{O}\left(\frac{1}{r^2}\right),\quad
  g_{t\varphi}=-\frac{2S_1\sin^2{\vartheta}}{r}+\mathcal{O}\left(\frac{1}{r^2}\right),
\end{equation}
we immediately conclude that the mass monopole $M_0$ vanishes, which is consistent with the vanishing Komar/ADM mass on each end, and the current dipole $S_1$ coincides with $J$. These are exactly the lowest GH moments. Let us also recall that the Thorne and GH moments are equivalent for stationary spacetimes \cite{Gursel1983GRG}. Since our metric is asymptotically flat on each end and the curvature scalars together with the $T_{\mu\nu}$ components decay sufficiently fast, we are in the regime where the GH formalism and its generalisation to non-vacuum stationary spacetimes apply \cite{Mayerson2023SP}. As a first step, we need to put our metric \eqref{metricell} into the Weyl--Lewis--Papapetrou (WLP) form. To do that, we notice that in a stationary axisymmetric spacetime with commuting Killing fields $\partial_t$ and $\partial_\varphi$, we can introduce the $2\times 2$ Gram matrix on the group orbits by $\lambda_{\mu\nu}=g(\xi_\mu,\xi_\nu)$. Following the standard uniqueness theorem reduction (see \cite{Heusler2010} and \cite{Chrusciel2012LRR}), we define the Weyl radius by $\rho^2 = -\det{\lambda_{\mu\nu}} = g^2_{t\varphi} - g_{tt} g_{\varphi\varphi}$. For our present metric, the Weyl radial coordinate becomes
\begin{eqnarray}\label{star}
  \rho(\ell,\chi)=\sqrt{(\ell^2+r_0^2)(1-\chi^2)}.
\end{eqnarray}
As expected, this is just $r\sin{\vartheta}$ if we recall that $r^2=\ell^2+r_0^2$ and $\chi=\cos{\vartheta}$. Now we need to pick a coordinate $z$ on the meridional $2$-space such that the metric on the $(\ell,\chi)$ submanifold becomes conformally flat in the $(\rho,z)$ coordinates. More precisely, we have
\begin{equation}
  d\sigma^2=d\ell^2+\frac{\ell^2+r_0^2}{1-\chi^2}d\chi^2.
\end{equation}
Let $z(\ell,\chi) = \ell\chi$. Then, a straightforward computation shows that
\begin{equation}
  d\rho^2+dz^2=\frac{\ell^2+r_0^2\chi^2}{\ell^2+r_0^2}d\sigma^2.
\end{equation}
Hence, we conclude that the meridional metric is conformally flat with $d\sigma^2 = \Lambda(\ell,\chi)(d\rho^2 + dz^2)$ and conformal factor $\Lambda(\ell,\chi) = (\ell^2 + r_0^2)/(\ell^2 + r_0^2\chi^2)$ with $\ell$ and $\chi$ understood as functions of $\rho$ and $z$. More precisely, we have
\begin{equation}\label{***}
  \ell^2=\frac{1}{2}\left(\rho^2+z^2-r_0^2+\sqrt{R_+ R_-}\right),\quad \chi=\frac{z}{\ell},\quad R_\pm=(\rho\pm r_0)^2+z^2.    
\end{equation}
Finally, the WLP form of our metric is \cite{Lewis1932PRSLA, Papapetrou1953AP, Soffel2019} 
\begin{equation}
  ds^2=-f(\rho,z)]\left[dt-\omega_{WLP}(\rho,z)d\varphi^2\right]^2+f^{-1}(\rho,z)\left[e^{2\gamma(\rho,z)}(d\rho^2+dz^2)+\rho^2d\varphi^2\right].    
\end{equation}
In this form, $g_{tt} = -f$, $g_{t\varphi} = f\omega_{WLP}$, $g_{\varphi\varphi} = f^{-1}\rho^2 - f\omega^2_{WLP}$. Moreover, we have
\begin{equation}
  f=1-\rho^2\omega^2(\ell),\quad
  \omega_{WLP}=-\frac{\rho^2\omega(\ell)}{1-\rho^2\omega^2(\ell)},\quad
  e^{2\gamma}=f\Lambda=\frac{(\ell^2+r_0^2)\left[1-\rho^2\omega^2(\ell)\right]}{\ell^2+r_0^2\chi^2}.
\end{equation}
Notice that along the symmetry axis, \emph{i.e.} $\rho = 0$ and $\chi = \pm 1$, we have for all $z$ 
\begin{equation}
  f(\rho=0,z) = 1, \quad \omega_{WLP}(\rho=0,z) = 0.
\end{equation}
For any stationary spacetime, the twist 1-form of the Killing vector field $\xi^\mu=\partial_t$ is defined by \cite{Wald1986}
\begin{equation}\label{*}
  \omega_\mu=e_{\mu\nu\alpha\beta}\xi^\nu\nabla^\alpha\xi^\beta
\end{equation}
with $e_{\mu\nu\alpha\beta}=\sqrt{-g}\epsilon_{\mu\nu\alpha\beta}$ the volume form and $\epsilon_{\mu\nu\alpha\beta}$ the Levi-Civita symbol. For our problem, we find
\begin{equation}\label{K}
  \omega_t=\omega_\varphi=0,\quad
  \omega_\rho=-\frac{f^2}{\rho}\partial_z\omega_{WLP},\quad
  \omega_z=\frac{f^2}{\rho}\partial_\rho\omega_{WLP}.
\end{equation}
Details of the derivation are provided in Appendix~\ref{Aa}. At this point, a remark is in order. Given a stationary spacetime with Killing vector $\xi^\mu = \partial_t$, we can define its norm $\lambda = -\xi^\mu\xi_\mu = f$ and the associated twist 1-form $\omega_\mu$ by \eqref{*}. If we take the curl of $\omega_\mu$, we get
\begin{equation}\label{1a}
  \partial_{[\mu}\omega_{\nu]}=-e_{\mu\nu\rho\sigma}\xi^\rho R^{\sigma}{}_\tau\xi^\tau.    
\end{equation}
In vacuum, $R_{\mu\nu}=0$ so $\partial_{[\mu}\omega_{\nu]}=0$. This signals that the twist 1-form is closed and hence, it is locally exact with $\omega_\mu = \nabla_\mu\Psi$ for some twist scalar $\Psi$. Then, one defines the Ernst potential $\mathcal{E} = f + i\Psi$ which satisfies the usual Ernst equation. In the presence of matter, as in our case, $R_{\mu\nu}\neq 0$ and the right-hand side of \eqref{1a} does not vanish. As a consequence, $\omega_\mu$ is not closed in general. Hence, we cannot simply set $\omega_\mu=\nabla_\mu\Psi$ for the twist 1-form we have already computed. However, \cite{Mayerson2023SP} proved that for any stationary solution of the Einstein field equations with arbitrary matter and under mild assumptions, one can build an improved twist 1-form $\omega_\mu^{I}$ whose curl vanishes, and hence, a global twist scalar exists for the twist 1-form $\omega_\mu+\omega_\mu^{I}$. For our spinning MT wormhole, the improved twist 1-form turns out to be
\begin{eqnarray}
  \omega_\chi^{I}&=&-\frac{27\pi^3 J^5}{4r_0^6}(1-\chi^2)+36\pi J^5(1-\chi^2)G(\ell),\\
  G(\ell)&=&-\frac{3}{4r_0^6}\arctan^2{\left(\frac{\ell}{r_0}\right)}+\frac{3\pi\ell^2-6r_0\ell+3\pi r_0^2}{4r_0^6(\ell^2+r_0^2)}\arctan{\left(\frac{\ell}{r_0}\right)}+\frac{3\ell(\pi\ell^2-r_0\ell+\pi r_0^2)}{4r_0^5(\ell^2+r_0^2)^2},\label{GL}\\
  \omega_t^{I}&=&\omega_\ell^{I}=\omega_\varphi^{I}=0.
\end{eqnarray}
For the derivation of the above result, we refer to Appendix~\ref{Ab}. Because of the prefactor $(1-\chi^2)$, we have $\omega^{I}_\chi(\ell,\chi=\pm 1)=0$. Hence, in the Weyl--Papapetrou coordinates $(\rho,z)$, the improved twist vector vanishes on the symmetry axis. This is why the twist scalar $\Psi$ along the rotational axis is completely determined by the twist $\omega_\mu$ alone. As a result, the Ernst potential on the axis, which we will use to construct the Fodor--Hoenselaers--Perj$\acute{\mbox{e}}$s (FHP) potential \cite{Fodor1989JMP}, is unaffected by the matter presence and all GH multipoles can be extracted in the usual manner. Let us consider the Ernst potential $\mathcal{E}=f+i\Psi$ with $\Psi$ satisfing $\partial_\mu\Psi=\omega_\mu$. Along the symmetry axis, let $\Psi=\Psi_{ax}(z)$. Then,
\begin{equation}\label{tilde}
  \frac{d\Psi_{ax}}{dz}=\omega_z(\rho=0,z)    
\end{equation}
together with the condition $\Psi_{ax}\to 0$ as $z\to\infty$. In order to solve \eqref{tilde}, we need the near-axis behaviour of $\omega_z$ defined by \eqref{K}. By means of \eqref{***} we find
\begin{equation}
    \ell(\rho,z)=z+\mathcal{O}(\rho^2),\quad
    1-\chi^2=\frac{\rho^2}{\ell^2+r_0^2}=\mathcal{O}(\rho^2),\quad
    f(\rho,z)=1+\mathcal{O}(\rho^2).
\end{equation}
This implies $\omega_{WLP}(\rho,z)=-\rho^2\omega(z)+\mathcal{O}(\rho^4)$ and differentiating at fixed $z$, we find $\partial_\rho\omega_{WLP}=-2\rho\omega(z)+\mathcal{O}(\rho^3)$, which replaced into the last equation in \eqref{K} gives 
\begin{equation}
  \omega_z(\rho,z)=-2\omega(z)+\mathcal{O}(\rho^2).
\end{equation}
Taking the axis limit $\rho\to 0$ yields $\omega_z(\rho=0,z)=-2\omega(z)$. Hence, we need to solve the differential equation
\begin{equation}
  \frac{d\Psi_{ax}}{dz}=-2\omega(z).
\end{equation}
The general solution reads
\begin{equation}
  \Psi_{ax}(z) = -\frac{6Jz}{r_0^3}\cot^{-1}{\left(\frac{z}{r_0}\right)} + c.
\end{equation}
Imposing the boundary condition $\Psi_{ax}\to 0$ as $z\to\infty$ fixes $c = 6J/r_0^2$ and we conclude that
\begin{equation}
  \Psi_{ax}(z)=\frac{6J}{r_0^2}-\frac{6Jz}{r_0^3}\cot^{-1}{\left(\frac{z}{r_0}\right)}=\frac{2J}{z^2}-\frac{6Jr_0^2}{5z^4}+\mathcal{O}\left(\frac{1}{z^6}\right). 
\end{equation}
Since on the axis $f = 1$, we conclude that the axial Ernst potential is
\begin{equation}
  \mathcal{E}_{ax}(z)=1+i\Psi_{ax}(z)=1+i\left[\frac{2J}{z^2}-\frac{6Jr_0^2}{5z^4}+\frac{6Jr_0^4}{7z^6}+\mathcal{O}\left(\frac{1}{z^8}\right)\right].  
\end{equation}
Following \cite{Fodor1989JMP} we can introduce a modified potential $\mathcal{V}=(1-\mathcal{E})/(1+\mathcal{E})$ in terms of the M\"{o}bius transform of the Ernst potential. Notice that in our problem $\mathcal{V}\to 0$ as $z\to\infty$. On the rotational axis, we find
\begin{equation}\label{alpha}
\mathcal{V}(z)=-\frac{i\Psi_{ax}(z)}{1+i\Psi_{ax}(z)}
=-i\frac{J}{z^2}+\left(\frac{3}{5}iJr_0^2-J^2\right)\frac{1}{z^4}
+\left(iJ^3-i\frac{3}{7}Jr_0^4+\frac{6}{5}J^2 r_0^2\right)\frac{1}{z^6}+\mathcal{O}\left(\frac{1}{z^8}\right).
\end{equation}
On the other hand, \cite{Fodor1989JMP} expands the axis potential in the Weyl $z$-coordinate as
\begin{equation}\label{beta}
  \mathcal{V}(\rho=0,z)=\sum_{n=0}^\infty\frac{m_n}{z^{n+1}},
\end{equation}
where the complex coefficients $m_n$ are the input data for the FHP multipole algorithm. Comparing \eqref{alpha} with \eqref{beta} gives
\begin{equation}
m_0=0,\quad
m_1=-iJ,\quad
m_2=0,\quad
m_3=-J^2+i\frac{3}{5}Jr_0^2,\quad
m_4=0,\quad
m_5=\frac{6}{5}J^2 r_0^2+i\left(J^3-\frac{3}{7}Jr_0^4\right)
\end{equation}
and so on. Furthermore, \cite{Fodor1989JMP} gives explicit algebraic relations between the $m_n$'s and the GH multipoles $P_n=M_m+iS_n$. However, \cite{Sotiriou2004CQG} provides the corrected FHP expressions for the gravitational multipoles $P_n$ in terms of the axis coefficients $m_n$. More precisely, we have
\begin{equation}
P_0=m_0,\quad
P_1=m_1,\quad
P_2=m_2,\quad
P_3=m_3,\quad
P_4=m_4-\frac{m_0}{7}M_{20},\quad
P_5=m_5+\frac{m_1}{21}M_{20}-\frac{m_0}{3}M_{30}
\end{equation}
with $M_{ij} = m_i m_j - m_{i-1}m_{j+1}$. Hence, the quadrupole and Octupole are
\begin{equation}
    P_2=0,\quad
    P_3=-J^2+\frac{3}{5}iJr_0^2.
\end{equation}
Moreover, we also have
\begin{equation}
  P_4 = 0, \quad
  P_5 = \frac{6}{5}J^2 r_0^2 + i\left(\frac{20}{21}J^3 - \frac{3}{7}Jr_0^4\right).
\end{equation}
Our computations reveal that, as expected, the spinning MT wormhole is massless in the GH sense ($M_0=0$), it has a pure spin dipole $S_1 = -J$ in our present choice of azimuthal orientation, no mass quadrupole $(M_2 = 0)$ and the first non-trivial higher multipole is the current octupole $S_3 = 3 Jr_0^2/5$ and the mass octupole $M_3 = -J^2$. In other words, our wormhole is a massless but spinning object whose first non-trivial structure appears at the octupole level, and it depends on the throat size $r_0$. This is radically different from Kerr, where any spin automatically drags along a mass $M$ and a nonzero quadrupole $M_2 = -J^2/M$. It is interesting to notice that the asymptotic multipole structure retains information about the wormhole throat.

\section{Conclusions and outlook}

In this work, we have constructed and analysed an exact spinning generalisation of the Morris--Thorne traversable wormhole supported by an anisotropic fluid. Starting from a Teo-type stationary ansatz with unit lapse, Morris--Thorne shape function $b(r) = r_0^2/r$, and a spherically symmetric throat, we solved the reduced Einstein equations in closed form and obtained an explicit frame–dragging function. The resulting metric describes a two–parameter family of asymptotically flat spacetimes labelled by the throat radius $r_0$ and the total angular momentum $J$, with curvature invariants that remain finite at the throat and reduce smoothly to the static Morris--Thorne/Ellis--Bronnikov configuration in the nonrotating limit. The matter content is modelled as an anisotropic fluid in a comoving orthonormal frame. The corresponding stress–energy tensor exhibits the expected reflection symmetry about the equatorial plane and violates all standard energy conditions, as required for traversable wormholes.

Rotation introduces qualitatively new features. We showed that there exists a critical spin $|J_c| = 2r_0^2/(3\pi)$ that controls the onset of an ergoregion. For $|J| < |J_c|$, no ergoregion is present, while at $|J| = |J_c|$, the ergosurface degenerates in the sense that it coincides with the throat. Finally, for $|J|>|J_c|$, a genuine ergoregion forms around the throat. At the same time, the lapse remains unity and the time coordinate defines a global temporal function, so that the spacetime is stably causal and does not admit closed timelike curves for any value of $J$. This provides an explicit example in which ergoregions and good causal behaviour coexist in a rotating wormhole geometry.

On the observational side, we studied the shadows of the spinning Morris–Thorne wormhole following the formalism of \cite{Deligianni2021PRD, Shaikh2018PRD}. The shadows we obtain are systematically smaller than those of a Kerr black hole with comparable parameters and exhibit a nontrivial dependence on the wormhole shape function, and hence on $r_0$. In particular, the rotation function inherits the shape–function dependence and imprints it on the shadow boundary. This contradicts naive claims that the shadow of a rotating wormhole is independent of the choice of shape function and instead shows that the shadow can, in principle, serve as a probe not only of the spin but also of the underlying throat geometry.

We further characterised the spacetime through its GH multipole moments, thereby providing a coordinate–independent description of the far field. The explicit computation shows that our solution is not an awkward coordinate patch of Kerr but a genuinely different stationary spacetime. In the GH sense, the configuration is massless, with vanishing mass monopole ($M_0=0$) but nonzero spin dipole ($S_1\neq 0$). To a distant observer, the leading field is therefore purely rotational. Moreover, the mass quadrupole moment vanishes identically, $M_2=0$, so that the first deviation from a pure spin–dipole field appears only at the octupole level. The first nontrivial higher moments are the current octupole $S_3$ and mass octupole $M_3$, followed by $P_5$, and both the imaginary part of $P_3$ and the real and imaginary parts of $P_5$ contain explicit factors of $r_0$. In other words, the multipole hierarchy is highly unconventional compared to Kerr and to most astrophysical configurations, where any spin is accompanied by a nonzero mass quadrupole. Already at the level of GH multipoles, the spinning MT wormhole carries two independent hairs $(J, r_0)$, and the moments explicitly retain memory of the throat scale. 

These properties suggest several directions for future work. First, GH multipoles are the natural input for post–Newtonian and effective one–body descriptions of orbital dynamics, as well as for extreme EMRI modelling. Our exact moments provide the necessary starting point to compute precession rates, periastron shifts, and ultimately gravitational waveforms in this background. In Kerr, specific relations among orbital frequency, periastron precession, and Lense--Thirring precession follow from the tight multipole structure. In the present spacetime, where ($M_2 = 0$) but higher spin–dependent multipoles are nonzero, these relations will be modified in a characteristic way. Second, in principle, sufficiently precise measurements of the external gravitational field—for example through EMRIs observed by future space–based detectors could place constraints on the throat radius $r_0$, even if the throat lies inside an ergoregion. Finally, the spinning Morris–Thorne wormhole provides a fully analytic, causally well–behaved benchmark for exploring how deviations from Kerr might manifest themselves in shadow imaging, strong–field lensing, and gravitational–wave astronomy. Further investigations along these lines are left for future work.

\appendix

\section{Derivation of the twist 1-form components}\label{Aa}

For the WLP form of our metric we have $g_{tt} = -f$, $g_{t\varphi} = f\omega_{WLP}$, $g_{\varphi\varphi} = f^{-1}\rho^2-f\omega^2_{WLP}$, and $g_{\rho\rho} = g_{zz} = f^{-1} e^{2\gamma}$ where all metric components depend on $\rho$, $z$. Consider the timelike Killing vector field $\xi^\mu=(1,0,0,0)$. Its covariant components are $\xi_\mu=g_{\mu\nu}\xi^\nu=g_{\mu t}$. Hence, we find $\xi_t=-f$, $\xi_\rho=\xi_z=0$, and $\xi_\varphi=f\omega_{WLP}$. To compute the contravariant components of the metric tensor, we need only consider the $2\times 2$ $(t,\varphi)$ block. A straightforward computation gives $g^{tt}=-f^{-1}+f\rho^{-2}\omega^2_{WLP}$, $g^{t\varphi}=f\rho^{-2}\omega_{WLP}$, $g^{\varphi\varphi}=f\rho^{-2}$, and $g^{\rho\rho}=g^{zz}=fe^{-2\gamma}$. Moreover, $g=-\rho^2 f^{-2}e^{4\gamma}$ and hence, $\sqrt{-g}=\rho f^{-1}e^{2\gamma}$. For the orientation we pick $e_{t\rho z\varphi}=\sqrt{-g}$. Recall that the twist 1-form of $\xi^\mu$ is given by \eqref{*}. For a Killing vector, it is convenient to rewrite \eqref{*} in terms of the exterior derivative of the 1-form $\xi_\mu$. Define $F_{\alpha\beta}=2\nabla_{[\alpha}\xi_{\beta]}$. Since we work in a torsion-free connection, we conclude that $\nabla_\alpha\xi_\beta-\nabla_\beta\xi_\alpha=\partial_\alpha\xi_\beta-\partial_\beta\xi_\alpha$, and therefore, $F_{\alpha\beta} = \partial_\alpha\xi_\beta - \partial_\beta\xi_\alpha = (d\xi)_{\alpha\beta}$. Notice that $F_{\alpha\beta}$ is antisymmetric by construction, and moreover, for a Killing field, $\nabla_\alpha\xi_\beta$ is purely antisymmetric, \emph{i.e.} $\nabla_{(\alpha}\xi_{\beta)} = 0$. This implies that $\nabla^\alpha\xi^\beta = F^{\alpha\beta}/2$. As a result, \eqref{*} can be rewritten as
\begin{equation}
  \omega_\mu=\frac{1}{2}e_{\mu\nu\alpha\beta}\xi^\nu F^{\alpha\beta}.    
\end{equation}
Since $\xi^\nu=\delta^\nu_t$, only $\nu=t$ contributes and we conclude that
\begin{equation}\label{II}
  \omega_\mu=\frac{1}{2}e_{\mu t\alpha\beta} F^{\alpha\beta}.    
\end{equation}
As a first step, we compute $F_{\mu\nu}$. The only non-zero derivatives are those with respect to $\rho$ and $z$. We find $F_{t\rho}=\partial_\rho f$, $F_{tz}=\partial_z f$, $F_{\rho\varphi}=\partial_\rho(f\omega_{WLP})$, and $F_{z\varphi}=\partial_z(f\omega_{WLP})$. By means of $F^{\alpha\beta}=g^{\alpha\mu}g^{\beta\nu}F_{\mu\nu}$, we obtain $F^{\rho\varphi}=f^3\rho^{-2}e^{-2\gamma}\partial_\rho\omega_{WLP}$ and $F^{z\varphi}=f^3\rho^{-2}e^{-2\gamma}\partial_z\omega_{WLP}$. Notice that all other $F^{\alpha\beta}$ are not needed in the present computation. Going back to \eqref{II}, we get $\omega_t=2^{-1}e_{tt\alpha\beta}F^{\alpha\beta}=0$ by antisymmetry of $e_{\mu\nu\alpha\beta}$. Moreover, we have
\begin{equation}
  \omega_\rho=\frac{1}{2}e_{\rho t\alpha\beta}F^{\alpha\beta} = \frac{1}{2}e_{\rho t z\varphi}F^{z\varphi}+\frac{1}{2}e_{\rho t\varphi z}F^{\varphi z} = e_{\rho t z\varphi}F^{z\varphi}=-e_{t\rho z\varphi}F^{z\varphi}=-\frac{f^2}{\rho}\partial_z\omega_{WLP}.
\end{equation}
Since $F^{\rho z} = g^{\rho\rho}g^{zz}F_{\rho z} = 0$, we immediately conclude that $\omega_\varphi = 0$. Finally, a computation similar to that for $\omega_\rho$ gives $\omega_z = f^2\rho^{-1}\partial_\rho\omega_{WLP}$.

\section{Computation of the improved twist 1-form}\label{Ab}

Let $\xi^\mu = \partial_t$ be a timelike Killing vector field. Define its norm as $\lambda = -\xi^\mu\xi_\mu$. The spatial metric, \emph{i.e.} the projector onto the 3-manifold $M$ of Killing orbits is \cite{Mayerson2023SP}
\begin{equation}
  h_{\mu\nu}=g_{\mu\nu}+\frac{\xi_\mu\xi_\nu}{\lambda}.
\end{equation}
As in \cite{Mayerson2023SP}, we define the matter current on $M$ as
\begin{equation}\label{A}
  V_\mu=8\pi h_{\mu}{}^\sigma T_{\sigma\nu}\xi^\nu=8\pi\left(T_{\mu\nu}\xi^\nu+\frac{1}{\lambda}\xi_\mu\xi^\sigma T_{\sigma\nu}\xi^\nu\right).    
\end{equation}
\cite{Mayerson2023SP} showed that such $V_\mu$ is divergence free and $\xi^\mu V_\mu=0$. Moreover, its Hodge dual is closed, that is $d(*V)=0$. As in \cite{Mayerson2023SP}, we define the following 2-form on $M$
\begin{equation}\label{B}
  W^{(2)}=i_{\xi}(*V),\quad W_{\mu\nu}=e_{\mu\nu\rho\sigma}\xi^\rho V^\sigma.    
\end{equation}
Under the assumption that there is no net radial momentum flux through any spatial 2-cycle, $W^{(2)}$ is not only closed but also exact, so there exists a 1-form $B^{(1)}$ on $M$ such that $W^{(2)}=dB^{(1)}$ with components $W_{\mu\nu} = \partial_{\mu}B^{(1)}_{\nu} - \partial_{\nu}B^{(1)}_{\mu}$. Then, the improved twist vector is defined by $\omega_\mu^I = 2B^{(1)}_\mu$ and satisfies $\partial_\mu\omega_\nu^{(I)} - \partial_\nu\omega_\mu^{(I)} = e_{\mu\nu\rho\sigma}\xi^\rho V^\sigma$. The total twist is $\omega_\mu^{tot} = \omega_\mu + \omega_\mu^{I}$, and it is curl-free, so we can define the twist scalar $\Psi$ by $\nabla_\mu\Psi = \omega_\mu^{tot}$ and the Ernst potential is $\mathcal{E} = f + i\Psi$. Let us rewrite our metric \eqref{metricell} as
\begin{equation}
  ds^2 = -\left[1-A(\ell,\chi)\omega^2(\ell)\right]dt^2+d\ell^2-2A(\ell,\chi)\omega(\ell)dtd\varphi + (\ell^2+r_0^2)\left[\frac{d\chi^2}{1-\chi^2}+(1-\chi^2)d\varphi^2\right]    
\end{equation}
with $\omega(\ell) = J\Omega(\ell)$ and $A(\ell,\chi) = (1 - \chi^2)(\ell^2 + r_0^2)$. Pick the timelike Killing vector $\xi^\mu=(1,0,0,0)$. Its covariant components are $\xi_t = -(1 - A\omega^2)$, $\xi_\varphi = - A\omega$, and $\xi_\ell = xi_\chi = 0$. The norm of $\xi^\mu$ is $\lambda = 1 - A\omega^2$. Moreover, $g = -(\ell^2 + r_0^2)$ and $\sqrt{-g} = \ell^2 + r_0^2$. The non-zero components of $T_{\mu\nu}$ are $T_{tt}=\widetilde{\rho}+A\omega^2 P_{\varphi}$, $T_{t\varphi}=-A\omega P_{\varphi}$, $T_{\varphi\varphi}=AP_{\varphi}$, $T_{\ell\ell}=P_r$, and $T_{\chi\chi}=(\ell^2+r_0^2)P_\chi/(1-\chi^2)$. Let us compute $V_\mu$ by means of \eqref{A}. Since $\xi^\nu=(1,0,0,0)$, we have
\begin{equation}
  V_\mu=8\pi\left(T_{\mu\nu}+\frac{1}{\lambda}\xi_\mu T_{tt}\right).
\end{equation}
Then, $V_t=V_\ell=V_\chi=0$ and $V_\varphi=-8\pi A\omega(\widetilde{\rho}+P_\varphi)/(1-A\omega^2)$. The contravariant components of the matter current on $M$ are
\begin{eqnarray}
  V^t&=&\frac{8\pi A\omega^2(\widetilde{\rho}+P_{\varphi})}{1-A\omega^2}=-\frac{18J^4A\omega^2(1-\chi^2)}{(1-A\omega^2)(\ell^2+r_0^2)^3},\\
  V^\ell&=&V^\chi=0,\\
  V^\varphi&=&-8\pi\omega(\widetilde{\rho}+P_{\varphi})=\frac{18J^4\omega(1-\chi^2)}{(\ell^2+r_0^2)^3}.
\end{eqnarray}
Let us now compute the 2-form given in \eqref{B}. In the present case, $g = -(\ell^2 + r_0^2)$. For the orientation, we choose $e_{t\ell\chi\varphi} = \ell^2 + r_0^2$. Moreover, $\xi^\nu = \delta^\nu_t$. Hence, we find $W_{\mu\nu}=e_{\mu\nu t\varphi}V^\varphi$. The only non zero components are $W_{\ell\chi}=-W_{\chi\ell}$ and
\begin{equation}
  W_{\ell\chi} = e_{\ell\chi t\varphi}V^\varphi = e_{t\ell\chi\varphi}V^\varphi = \frac{18J^4\omega(1-\chi^2)}{(\ell^2+r_0^2)^2}.    
\end{equation}
Hence, $W^{(2)}$ is represented by the following 2-form on the $(\ell,\chi)$-plane
\begin{equation}\label{C}
  W^{(2)} = W_{\ell\chi}d\ell\wedge d\chi = \frac{18J^4\omega(1-\chi^2)}{(\ell^2 + r_0^2)^2}d\ell\wedge d\chi.
\end{equation}
Let us solve $W^{(2)} = dB^{(1)}$. Since $W^{(2)}$ has only a $d\ell\wedge d\chi$ component and no $t$ or $\varphi$ dependence, the simplest ansatz consistent with axisymmetry and stationarity is $B^{(1)} = B(\ell,\chi)d\chi$ with $B(\ell,\chi)$ a scalar function. Then $dB^{(1)} = \partial_\ell B(\ell,\chi)d\ell\wedge d\chi$ which matched to \eqref{C} gives the following partial differential equation
\begin{equation}\label{D}
  \partial_\ell B(\ell,\chi)=\frac{18J^4\omega(1-\chi^2)}{(\ell^2+r_0^2)^2}.
\end{equation}
We impose the boundary condition $B(\ell,\chi)\to 0$ as $\ell\to+\infty$. Then, integrating \eqref{D} yields
\begin{equation}
  B(\ell,\chi) = F(\chi) + 18\pi J^5(1 - \chi^2)G(\ell),\quad
  G(\ell) = \int\frac{\Omega(\ell)d\ell}{(\ell^2 + r_0^2)^2}.
\end{equation}
The integral defining $G(\ell)$ can be computed exactly, and it is given by \eqref{GL}. The unknown integration function $F(\chi)$ is fixed by imposing the boundary condition $B(\ell,\chi) \to 0$ as $\ell \to +\infty$. More precisely, we find $F(\chi)=-27\pi^3 J^5(1-\chi^2)/(8r_0^6)$. Finally, the improved twist vector is $\omega_\mu^I=2B^{(1)}_\mu$ but $B^{(1)}$ has only a $\chi$-component and hence, we conclude that $\omega^I_t = \omega^I_\ell = \omega^I_\varphi = 0$ and $\omega^I_\chi = 2 B(\ell,\chi)$. 

\bibliography{QNMS}

\end{document}